\documentclass[twocolumn,aps,pre,amsmath,reprint,longbibliography,nofootinbib,superscriptaddress]{revtex4-2}
\usepackage{amsmath,amssymb}
\usepackage{graphicx}
\usepackage{amsthm}
\usepackage[colorlinks=true, bookmarks=true,
bookmarksnumbered=true, bookmarkstype=toc, linkcolor=magenta,
urlcolor=cyan, citecolor=cyan]{hyperref}
\newtheorem{proposition}{Proposition}
\newtheorem*{contrapositive*}{Contrapositive of Proposition 1}
\newtheorem*{observation*}{Observation}
\begin{document}
\title{
Temperature chaos as a logical consequence of the reentrant transition in spin glasses
}
\author{Hidetoshi Nishimori}
\affiliation{Institute of Integrated Research, Institute of Science Tokyo, Nagatsuta-cho, Midori-ku, Yokohama 226-8501, Japan}
\affiliation{Graduate School of Information Sciences, Tohoku University, Sendai 980-8579, Japan}
\affiliation{RIKEN Interdisciplinary Theoretical and Mathematical Sciences (iTHEMS), Wako, Saitama 351-0198, Japan}
\author{Masayuki Ohzeki}
\affiliation{Graduate School of Information Sciences, Tohoku University, Sendai 980-8579, Japan}
\affiliation{Department of Physics, Institute of Science Tokyo, Tokyo 152-8551, Japan}
\affiliation{Sigma-i Co., Ltd., Tokyo 108-0075, Japan}
\author{Manaka Okuyama}
\affiliation{Graduate School of Information Sciences, Tohoku University, Sendai 980-8579, Japan}
\date{\today}
\begin{abstract} 
Temperature chaos is a striking phenomenon in spin glasses, where even slight changes in temperature lead to a complete reconfiguration of the spin state. Another intriguing effect is the reentrant transition, in which lowering the temperature drives the system from a ferromagnetic phase into a less ordered spin-glass or paramagnetic phase. In the present paper, we reveal an unexpected connection between these seemingly unrelated phenomena in the finite-dimensional Edwards-Anderson model of spin glasses by introducing a generalized formulation that incorporates correlations among disorder variables. Assuming the existence of a spin glass phase at finite temperature, we establish that temperature chaos arises as a logical consequence of reentrance in the Edwards-Anderson model. Our findings uncover a previously hidden mathematical structure relating reentrance and temperature chaos, offering a new perspective on the physics of spin glasses beyond the mean-field theory.
\end{abstract}
\maketitle

\section{Introduction}
\label{sec:introduction}
Statistical-mechanical studies of spin glasses have uncovered a number of striking phenomena, including replica symmetry breaking in the mean-field Sherrington–Kirkpatrick model \cite{Sherrington1975,Parisi1980,Talagrand2006}, as well as deep connections with fields beyond traditional physics, such as computer science \cite{Mezard2009}, inference problems \cite{Zdeborova2016}, and a number of other domains \cite{Nishimori2001,Charbonneau2023}.

Many of these theoretical developments concern mean-field-type models. By contrast, much less is known on firm theoretical grounds about the finite-dimensional Edwards–Anderson model \cite{Edwards1975}, especially in two and three dimensions, which are most relevant to experiments \cite{Mydosh1993}. For these cases, numerical simulations remain the primary investigative tool.

Numerical studies suggest the existence of two counterintuitive phenomena in the finite-dimensional Edwards–Anderson model, reentrant transitions and temperature chaos. A reentrant transition, where the lower-temperature phase is less ordered than the higher-temperature phase, has been observed in two, three, and four dimensions, near the boundary between the ferromagnetic and non-ferromagnetic (spin glass or paramagnetic) phases \cite{Nobre2001,Wang2003,Amoruso2004,Hasenbusch2007,Toldin2009,Ceccarelli2011,Thomas2011,Liu2025}.
Another notable phenomenon is temperature chaos, in which small changes in temperature within the spin glass phase lead to a complete reorganization of the spin configuration \cite{Bray1987,Banavar1987,Fisher1986,Fisher1988,Kondor1989,Ney-Nifle-1997,Ney-Nifle1998,Parisi2010,Mathieu2001,Bouchaud2001,Aspelmeier2002,Rizzo2003,Houdayer2004,Katzgraber2007b,Fernandez-2013,Wang2015,Billoire2018,Baity-Jesi2021}.
It is worth noting for later discussions that temperature chaos has been studied predominantly in the case with symmetric distributions of disorder, where positive and negative interactions occur with equal probability, far from the region of the phase diagram where reentrant transitions have been observed. As a result, little attention has been given to the possibility of a connection between the two.

One of the notable recent developments in spin glass theory is the discovery of the significant impact of spatial correlations in disorder variables \cite{Nishimori2024,Nishimori2025,Braunstein2025}. Spatially correlated disorder has been shown to induce anomalous behaviors, including straight, non-reentrant, phase boundaries between the ferromagnetic and non-ferromagnetic phases, under the condition that temperature chaos is absent. If temperature chaos is present, the ferromagnetic phase is confined to a single line, the Nishimori line (NL) \cite{Nishimori1981,Nishimori1980,Nishimori1981thesis,Nishimori2001}, and is surrounded by the spin glass phase in the model with strong correlations in disorder variables. Furthermore, assuming replica symmetry breaking in the Edwards–Anderson model, the distribution of the magnetization on the NL exactly matches the distribution of the replica overlap in the spin glass phase, with support on a finite interval. This result implies that the magnetization fluctuates from one measurement to another or from one disorder realization to another, which is an unusual property for a macroscopic quantity. However, these surprising conclusions pertain specifically to models with strongly correlated disorder and cannot be directly extended to the standard Edwards–Anderson model with uncorrelated disorder.

In the present paper, we extend the theoretical framework for correlated disorder proposed in Ref.~\cite{Nishimori2024} by introducing an additional parameter that controls the level of frustration. Analysis of the resulting model reveals an unexpected relation between temperature chaos and the reentrant transition in the Edwards–Anderson model with uncorrelated disorder. Specifically, we demonstrate that, under the assumption of the existence of a spin glass phase, the phase boundary between the ferromagnetic and spin glass phases is non-reentrant if temperature chaos is absent. The contrapositive of this statement is that temperature chaos exists if the phase boundary is reentrant. This establishes a highly nontrivial mathematical relationship between two seemingly unrelated physical phenomena observed in different regions of the phase diagram in the Edwards–Anderson model, which is derived through the analysis of a model with correlations in disorder variables beyond the Edwards-Anderson model.

The correlated disorder model introduced in this work is closely related to the model proposed by Kitatani \cite{Kitatani1992}, who used it to argue against the existence of a reentrant transition in the Edwards–Anderson model. Our model can be viewed as a reformulation of his model, by incorporating a more natural parametrization of the control variables. This formulation allows for a more direct and modern analysis, leading to conclusions that do not necessarily align with Kitatani’s.

The structure of this paper is as follows. We begin by introducing a correlated disorder model and show how it reduces to known models in several special cases in Sec.~\ref{sec:formulation}. Then, in Sec.~\ref{sec:main}, we first analyze several limiting cases to construct 2D cross sections of a 3D phase diagram.\footnote{To avoid confusion, we write ``3D" for the phase diagram with three axes and ``three dimensions" for the spatial dimensionality of the lattice.} In the later part of Sec.~\ref{sec:main}, we integrate these results to establish a logical relation between reentrance and temperature chaos from the 3D phase diagram, under the assumption of the existence of a spin glass phase. We conclude the paper in the final section.

\section{Formulation of the problem}
\label{sec:formulation}

Let us first introduce the model and define several physical quantities that are central to our discussion.

\subsection{Problem Definition}

We consider the $\pm J$ Ising spin glass with the dimensionless Hamiltonian
\begin{align}
    H = -\beta \sum_{\langle ij\rangle} \tau_{ij} S_i S_j \quad (S_i = \pm 1),
    \label{eq:Hamiltonian}
\end{align}
where $\beta$ is the inverse temperature (coupling constant), $\tau_{ij}(=\pm 1)$ denotes the quenched disorder variable associated with the bond $\langle ij \rangle$, and $S_i$ is the Ising spin at site $i$. The summation runs over all interacting spin pairs on a given lattice, with no restriction on the dimensionality or the structure of the lattice. Consequently, our theory applies also to the all-to-all interacting, infinite-range, Sherrington-Kirkpatrick model
\footnote{The parameters should be rescaled by appropriate powers of the system size for the Sherrington-Kirkpatrick model.}
in addition to the finite-dimensional Edwards-Anderson model if we choose the parameters appropriately as described below.

The configuration of disorder variables $\tau = \{\tau_{ij}\}$ is assumed to follow the probability distribution
\begin{align}
    P(\tau) = \frac{1}{A} \, Z_{\tau}(\gamma) \, \frac{e^{\beta_p \sum_{\langle ij\rangle} \tau_{ij}}}{Z_{\tau}(\beta_p)},
    \label{eq:P_correlated}
\end{align}
where $Z_{\tau}(\gamma)$ is the partition function of the Ising model for a given disorder realization $\tau$,
\begin{align}
    Z_{\tau}(\gamma) = \sum_S e^{\gamma \sum_{\langle ij\rangle} \tau_{ij} S_i S_j},
\end{align}
and $A$ is the normalization constant. The latter can be computed explicitly using the gauge transformation $\tau_{ij} \to \tau_{ij} \sigma_i \sigma_j$ with $\sigma_i = \pm 1$,
\begin{align}
    A &= \sum_{\tau} Z_{\tau}(\gamma) \frac{e^{\beta_p \sum_{\langle ij\rangle} \tau_{ij}}}{Z_{\tau}(\beta_p)} \nonumber \\
      &= \frac{1}{2^N} \sum_{\tau} Z_{\tau}(\gamma) \frac{\sum_{\sigma} e^{\beta_p \sum_{\langle ij\rangle} \tau_{ij} \sigma_i \sigma_j}}{Z_{\tau}(\beta_p)} \nonumber \\
      &= \frac{1}{2^N} \sum_S \sum_{\tau} e^{\gamma \sum_{\langle ij\rangle} \tau_{ij} S_i S_j} \nonumber \\
      &= (2\cosh \gamma)^{N_B},
    \label{eq:Ap}
\end{align}
where $N_B$ is the total number of bonds (i.e., interacting spin pairs).

The distribution in Eq.~\eqref{eq:P_correlated} describes correlated disorder, since it does not generally factorize into a product of independent distributions:
\begin{align}
    P(\tau) \ne \prod_{\langle ij \rangle} p(\tau_{ij})
\end{align}
for any function $p(\cdot)$, unless $\gamma = \beta_p$.

\subsection{Reduction to Known Models}
\label{subsub:reduction}
The distribution $P(\tau)$ in Eq.~(\ref{eq:P_correlated}) depends on two parameters, $\gamma$ and $\beta_p$, and reduces to the standard Edwards--Anderson model when $\gamma = \beta_p$:
\begin{align}
     P(\tau)\big|_{\gamma = \beta_p} = \frac{e^{\beta_p \sum \tau_{ij}}}{(2\cosh \beta_p)^{N_B}},
    \label{eq:P_EA}
\end{align}
in which the disorder variables $\tau_{ij}$ are independent and identically distributed (i.i.d.), and thus spatially uncorrelated.

The distribution $P(\tau)$ generalizes the correlated-disorder model discussed in Ref.~\cite{Nishimori2024}, corresponding to the special case $\gamma = 0$:
\begin{align}
    P(\tau)\big|_{\gamma = 0} = \frac{2^N}{2^{N_B}} \, \frac{e^{\beta_p \sum \tau_{ij}}}{Z_{\tau}(\beta_p)}.
    \label{eq:P_gamma0}
\end{align}

The model proposed by Kitatani \cite{Kitatani1992} can be recovered by identifying $\gamma \to K_p$ and $\beta_p \to K_p + a$. As will be shown in the following sections, our parametrization in Eq.~\eqref{eq:P_correlated} facilitates a more direct and natural analysis via the 3D phase diagram with axes $(\gamma, \beta_p, 1/\beta)$. A reinterpretation of Kitatani’s argument will be presented later in this paper.

Our distribution function of Eq.~(\ref{eq:P_correlated}) unifies these cases and provides a new perspective, especially on the most important case of the Edwards-Anderson model.

It is helpful to clarify the roles of the partition functions $Z_{\tau}(\gamma)$ and $Z_{\tau}(\beta_p)$ in the probability distribution $P(\tau)$ defined in Eq.~(\ref{eq:P_correlated}), a feature absent in the Edwards-Anderson model, Eq.~(\ref{eq:P_EA}).

Let us first focus on the term involving $\beta_p$, namely $e^{\beta_p\sum \tau_{ij}} / Z_{\tau}(\beta_p)$. When frustration is weaker, the free energy
\begin{align}
    \beta_p F_{\tau}(\beta_p) = -\ln Z_{\tau}(\beta_p)
\end{align}
tends to be lower, corresponding to a larger value of $Z_{\tau}(\beta_p)$, as systems with less frustration are typically more stable than those with stronger frustration. Indeed, in the limit $\beta_p \to \infty$, only frustration-free configurations of the disorder variables $\tau$ contribute significantly,
\begin{align}
    Z_{\tau}(\beta_p)\to
    \begin{cases}
        e^{N_B\beta_p} \qquad {\rm if~} \tau_{ij} = \xi_i\xi_j~ {\rm for~some~} \{\xi_i = \pm 1\}_i, \\
        e^{(N_B - a)\beta_p} \quad (\exists~a > 0) \qquad {\rm otherwise},
    \end{cases}
\end{align}
since frustration-free configurations possess the lowest possible ground-state energy, equal to that of the pure ferromagnetic system. This dominance of frustration-free configurations is expected to persist even at finite $\beta_p$. Accordingly, the factor $1/Z_{\tau}(\beta_p)$ increases the relative weight of disorder configurations with stronger frustration compared to the Edwards-Anderson model, where this factor is absent. We also note that, despite this bias toward stronger frustration, the fully ferromagnetic spin configuration remains one of the ground states generated by disorder configurations sampled from the distribution $e^{\beta_p\sum \tau_{ij}} / Z_{\tau}(\beta_p)$, as proven in the Appendix of Ref.~\cite{Nishimori2024}.

The additional factor $Z_{\tau}(\gamma)$ in Eq.~(\ref{eq:P_correlated}) partially compensates for the denominator $1/Z_{\tau}(\beta_p)$, as it appears in the numerator and enhances the probability of configurations with weaker frustration. As a result, the parameter $\gamma$ allows for more flexible control over the system properties compared to the model discussed in Ref.~\cite{Nishimori2024}, where this degree of freedom is absent. As we will show below, this added flexibility leads to the identification of a previously unrecognized relationship between temperature chaos and reentrance.

\subsection{Distribution functions of physical quantities}
The distribution function of the magnetization $P_1(x|\beta,\beta_p,\gamma)$ and the distribution function of the overlap $P_2(x|\beta_1,\beta_2,\gamma)$ of two replicas $\{S_i^{(1)}\}$ and $\{S_i^{(2)}\}$ with the same set of disorder $\{\tau\}$ are defined as follows:
\begin{widetext}
\begin{align}
    P_1(x|\beta,\beta_p,\gamma)
    &=\frac{1}{A}\sum_{\tau}Z_{\tau}(\gamma)\frac{e^{\beta_p \sum \tau_{ij}}}{Z_{\tau}(\beta_p)}
   \,
    \frac{\sum_S \delta\big(x-\frac{1}{N}\sum_iS_i\big)e^{\beta \sum \tau_{ij}S_iS_j}}{Z_{\tau}(\beta)}\nonumber\\
   &=\frac{1}{2^N A}\sum_{\tau}Z_{\tau}(\gamma)
    \frac{\sum_{S,\sigma} \delta\big(x-\frac{1}{N}\sum_i S_i \sigma_i\big)e^{\beta_p \sum \tau_{ij}\sigma_i \sigma_j} e^{\beta \sum \tau_{ij}S_i S_j}}{Z_{\tau}(\beta_p)Z_{\tau}(\beta)}\label{eq:p1def}\\
    P_2(x|\beta_1,\beta_2,\gamma)
    &=\frac{1}{A}\sum_{\tau}Z_{\tau}(\gamma)\frac{e^{\beta_p \sum \tau_{ij}}}{Z_{\tau}(\beta_p)}
    \,\frac{\sum_{S^{(1,2)}} \delta\big(x-\frac{1}{N}\sum_iS_i^{(1)}S_i^{(2)}\big)e^{\beta_1 \sum \tau_{ij}S_i^{(1)}S_j^{(1)}}e^{\beta_2 \sum \tau_{ij}S_i^{(2)}S_j^{(2)}}}{Z_{\tau}(\beta_1)Z_{\tau}(\beta_2)}\nonumber\\
    &=\frac{1}{2^N A}\sum_{\tau}Z_{\tau}(\gamma)
    \frac{\sum_{S^{(1,2)}} \delta\big(x-\frac{1}{N}\sum_iS_i^{(1)}S_i^{(2)}\big)e^{\beta_1 \sum \tau_{ij}S_i^{(1)}S_j^{(1)}}e^{\beta_2 \sum \tau_{ij}S_i^{(2)}S_j^{(2)}}}{Z_{\tau}(\beta_1)Z_{\tau}(\beta_2)}.
    \label{eq:P2}
\end{align}
\end{widetext}
In both equations, we have applied the gauge transformation $\tau_{ij} \to \tau_{ij} \sigma_i \sigma_j$ and $S_i \to S_i \sigma_i$ to move from the first expression to the second.

It is useful to note that $P_2(x|\beta_1, \beta_2, \gamma)$ is independent of $\beta_p$. This holds for any gauge-invariant quantity as can be verified by applying the same gauge transformation to move from the first to the second line of Eq.~(\ref{eq:P2}). An important consequence of this observation is that $P_2(x|\beta_1, \beta_2, \gamma)$ can be interpreted as the replica overlap for the Edwards-Anderson model defined in Eq.~(\ref{eq:P_EA}), with the disorder parameter $\beta_p$ replaced by $\gamma$. The same holds for any gauge-invariant quantity including the spin glass order parameter. This property will play a central role in the discussion that follows.

The magnetization and the spin glass order parameter are the average of $x$ by these distribution functions,
\begin{align}
    m(\beta,\beta_p,\gamma)&=\int_0^1 P_1(x|\beta,\beta_p,\gamma)\,xdx\\
    q(\beta_1,\beta_2,\gamma)&=\int_0^1 P_2(x|\beta_1,\beta_2,\gamma)\,xdx,
\end{align}
where the interval of integration is restricted to $x \ge 0$ to avoid trivial vanishing by the $\mathbb{Z}_2$ symmetry.

It is easy to see from the above equations that $P_1$ and $P_2$ are equal to each other,
\begin{align}
P_1(x|\beta,\beta_p,\gamma)=P_2(x|\beta_1=\beta,\beta_2=\beta_p,\gamma). \label{eq:p1_p2}
\end{align}
This is a remarkable relation because it connects the property of the present model on the left-hand side with that of the Edwards-Anderson model on the right-hand side, as mentioned in a previous paragraph.

Another useful relation is that $P_1(x|\beta,\beta_p,\gamma)$ is invariant under the exchange of $\beta$ and $\beta_p$,
\begin{align}
P_1(x|\beta,\beta_p,\gamma)=P_1(x|\beta_p,\beta,\gamma).
    \label{eq:P1}
\end{align}
These equations are central to the following discussions as was the case in Ref.~\cite{Nishimori2024}.

\section{Structure of the phase diagram}
\label{sec:main}
This is the main section of the paper, where we investigate the structure of the 3D phase diagram drawn with axes $\gamma, \beta_p$, and $1/\beta$. To this end, it is useful to start from the analyses of a few limiting cases, which provide 2D cross sections of the 3D phase diagram. 

Readers may find it useful to refer to the final results presented as the 3D phase diagrams in Figs.~\ref{fig:3D_NoTC_NoReent} and ~\ref{fig:3D_TC_Reent} even before they are cited explicitly in the main text. The former Fig.~\ref{fig:3D_NoTC_NoReent} is for the case without temperature chaos and reentrance, and the latter Fig.~\ref{fig:3D_TC_Reent} is for the case with temperature chaos and reentrance. Their 2D cross sections are discussed below to construct the final 3D phase diagrams.

\subsection{$\beta_p=0$}
\label{subsec:beta_p=0}
When $\beta_p=0$, the probability distribution of disorder is gauge invariant,
\begin{align}
    P(\tau)=\frac{Z_{\tau}(\gamma)}{2^N(2\cosh \gamma)^{N_B}}.\label{eq:P_Z}
\end{align}
In this case, the average of any gauge-invariant physical quantity is equal to the average for the Edwards-Anderson model with the disorder parameter value $\gamma$. For example, the energy is
\begin{align}
    E(\beta,0,\gamma)&=-\frac{1}{2^N(2\cosh\gamma)^{N_B}}\sum_{\tau}Z_{\tau}(\gamma)\frac{\partial_{\beta}Z_{\tau}(\beta)}{Z_{\tau}(\beta)}\nonumber\\
    &=-\frac{1}{(2\cosh \gamma)^{N_B}}\sum_{\tau}e^{\gamma \sum\tau_{ij}}\frac{\partial_{\beta}Z_{\tau}(\beta)}{Z_{\tau}(\beta)}.
\end{align}
This expression can be evaluated explicitly under the NL condition $\gamma=\beta$ as in the Edwards-Anderson model, with the result \cite{Nishimori2001,Nishimori1980,Nishimori1981,Nishimori1981thesis}
\begin{align}
    E(\beta,0,\beta)=-N_B\tanh \beta.
    \label{eq:exact_energy}
\end{align}

By contrast, the average of any gauge non-invariant quantity vanishes. For instance, the magnetization is, under fixed boundary conditions to avoid trivial vanishing,
\begin{align}
    &m(\beta,0,\gamma)\nonumber\\
    &=\frac{1}{2^{N}(2\cosh \gamma)^{N_B}}\sum_{\tau}Z_{\tau}(\gamma)\frac{\sum_S S_i\, e^{\beta\sum\tau_{ij}S_iS_j}}{Z_{\tau}(\beta)}\nonumber\\
        &=\frac{1}{2^{2N}(2\cosh \gamma)^{N_B}}\sum_{\tau}Z_{\tau}(\gamma)\frac{\sum_{\sigma}\sum_S S_i\sigma_i\, e^{\beta\sum\tau_{ij}S_iS_j}}{Z_{\tau}(\beta)}\nonumber\\
        &=0.
\end{align}

\begin{figure}[ht]
\centering
\includegraphics[width=60mm]{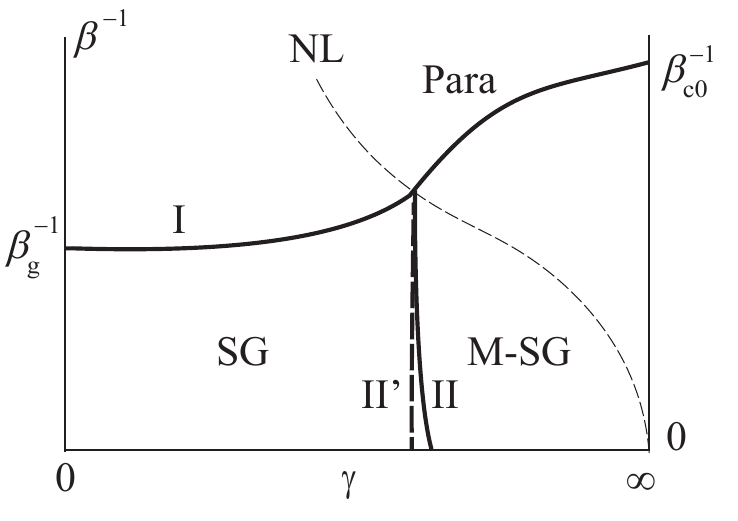}
\caption{Schematic phase diagram for $\beta_p = 0$, showing a cross section of the 3D phase diagram in Figs.~\ref{fig:3D_NoTC_NoReent} and \ref{fig:3D_TC_Reent}. The phase boundaries have the same shape as in the Edwards-Anderson model on the same lattice, with $\beta_p$ replaced by $\gamma$. The ferromagnetic phase of the Edwards-Anderson model is replaced by a Mattis-like spin glass phase (M-SG) with vanishing magnetization ($m = 0$). Reentrance is assumed along boundary II, but not along II'. NL denotes the Nishimori line ($\beta = \gamma$). Boundary I indicates the transition between paramagnetic and non-paramagnetic phases. ${\beta_{\rm c0}}^{-1}$ is the critical temperature of the pure Ising model, which is recovered in the limit $\gamma \to \infty$ for gauge-invariant quantities. ${\beta_{\rm g}}^{-1}$ is the spin glass transition temperature of the Edwards-Anderson model recovered at $\gamma=0$.
}
\label{fig:beta_p=0}
\end{figure}

As a consequence of these observations, the cross section of the 3D phase diagram at $\beta_p=0$ has the same structure as in the Edwards-Anderson model having $\gamma$ as the parameter to control the disorder, as long as gauge invariant quantities such as the spin glass order parameter are concerned. Consequently, the model has the same phase boundaries between the paramagnetic and spin glass/ferromagnetic phases as in the Edwards-Anderson model. See Fig.~\ref{fig:beta_p=0}.

It is important to remember that the ferromagnetic phase in the Edwards-Anderson model is replaced by the spin glass phase because $m(\beta,0,\gamma)=0$. Nevertheless, the boundary line between the ferromagnetic and spin glass phases, denoted II or II' in Fig.~\ref{fig:beta_p=0}, is very likely to keep existing in the present model because the spin glass order parameter and other gauge-invariant physical quantities such as the free energy are likely to have a singularity across the ferromagnetic-spin glass boundary in the Edwards-Anderson model, which is a property to be shared by the present model.

In the phase diagram of Fig. \ref{fig:beta_p=0}, we have assumed the existence of a spin glass phase in the Edwards-Anderson model on the same lattice, which is believed to be the case in three and higher dimensions, see e.g., Refs.~\cite{Hasenbusch2008,AltieriBaityJesi2023,Dahlberg2024}.

Notice that the ferromagnetic phase in the Edwards-Anderson model is replaced by the Mattis-like spin glass phase (M-SG). In this phase, the magnetization vanishes and the spin glass order parameter is finite. The distribution function of the overlap $P_2(x|\beta_1,\beta_2,\gamma)$ is gauge invariant and therefore has the same functional form as in the Edwards-Anderson model.  In the latter model, this region is in the ferromagnetic phase, and $P_2(x|\beta_1,\beta_2,\gamma)$ has two delta peaks when both $\beta_1$ and $\beta_2$ lie in this ferromagnetic phase,
\begin{align}
    &P_2(x|\beta_1,\beta_2,\gamma)\nonumber\\
    &=
    \displaystyle
    \begin{cases}
        \frac{1}{2}\delta(x-q)+\frac{1}{2}\delta(x+q)\quad&(\beta_1=\beta_2)\\
         \frac{1}{2}\delta(x-q_{12})+ \frac{1}{2}\delta(x+q_{12})\quad&(\beta_1\ne\beta_2)
    \end{cases},
    \label{eq:P2_EA}
\end{align}
where $q_{12}$ is the overlap of ferromagnetically-ordered spin states at two temperatures. Therefore, the Mattis-like spin glass phase in the present model has no replica symmetry breaking or temperature chaos, which are characterized by a non-trivial functional form of  $P_2(x|\beta_1,\beta_2,\gamma)$ for $\beta_1=\beta_2$ and the single delta function $\delta(x)$ for $\beta_1\ne\beta_2$. The reason for the name Mattis-like spin glass phase will be explained in Sec.~\ref{sub:mattis}. 

We point out that we have assumed the existence of the ferromagnetic phase in the Edwards-Anderson model for $\gamma$ above a threshold value.  This fact has so far been proven rigorously only in a restricted part of the phase diagram \cite{Nishimori1981thesis,Horiguchi1982} (\cite{garban2022continuous}).

It is worth noting that the Ising spin glass with the probability distribution proportional to the partition function in Eq.~(\ref{eq:P_Z}) has been experimentally implemented on a quantum computer under the NL condition $\gamma = \beta$, and the exact energy expression of Eq.~(\ref{eq:exact_energy}) has been successfully confirmed \cite{Chen2025,Zhu2023}. This is the first experiment to test this exact solution.

\subsection{Cross section at fixed $\beta_p$}
\label{subsec:beta_p_fixed}
Let us next fix $\beta_p$ to a finite value, not necessarily $\beta_p=0$. Since the average of any gauge-invariant quantity has no dependence on $\beta_p$, for example $P_2(x|\beta_1, \beta_2, \gamma)$ and the spin glass order parameter $q$, the structure of the phase diagram in Fig.~\ref{fig:beta_p=0} remains unchanged for any fixed value of $\beta_p$ as long as gauge-invariant quantities are concerned.

Notice, however, that the vanishing of the magnetization holds true only for small values of $\beta_p$ because
\begin{align}
    &m(\beta,\beta_p,\gamma)=\frac{1}{A}\sum_{\tau}Z_{\tau}(\gamma)\frac{e^{\beta_p\sum \tau_{ij}}}{Z_{\tau}(\beta_p)}\frac{\sum_S S_i\, e^{\beta\sum\tau_{ij}S_iS_j}}{Z_{\tau}(\beta)}\nonumber\\
        &=\frac{1}{2^N A}\sum_{\tau}Z_{\tau}(\gamma)
        \frac{\sum_{\sigma}\sigma_i\,e^{\beta_p\sum \tau_{ij}\sigma_i\sigma_j}\sum_{S}S_i\,e^{\beta\sum \tau_{ij} S_iS_j}}{Z_{\tau}(\beta_p)Z_{\tau}(\beta)},
\end{align}
which may become finite for larger $\beta_p$. Accordingly, the Mattis-like spin glass phase will be replaced by the ferromagnetic phase for large $\beta_p$.

\subsection{$\gamma\to\infty$}
\label{sub:mattis}
The configurations of disorder variables $\tau$ surviving in the limit $\gamma\to\infty$ are those without frustration $\tau_{ij}=\xi_i\xi_j~(\xi_i=\pm 1)$, i.e., the Mattis model \cite{Mattis1976}, as discussed in Sec.~\ref{subsub:reduction},
\begin{align}
    Z_{\tau}(\gamma)\to
    \begin{cases}
        e^{N_B\gamma} \qquad {\rm if~}\{\tau_{ij}=\xi_i\xi_j\}~ {\rm for~some~} \{\xi_i\}_i\\
        e^{(N_B-a)\gamma}~ (\exists ~a>0) \quad {\rm otherwise}.
    \end{cases}
\end{align}
Then, only those non-frustrated configurations remain in the sum over $\tau$ for the average of a function $f(\tau)$,
\begin{align}
    \sum_{\tau}P(\tau) f(\tau)\to \sum_{\xi}\frac{e^{\beta_p \sum \xi_i\xi_j}}{Z_{\rm I}(\beta_p)} f(\xi),
\end{align}
where $Z_{\rm I}(\beta_p)$ is the partition function of the pure ferromagnetic Ising model coming from $Z_{\tau}(\beta_p)$ in Eq.~(\ref{eq:P_correlated}), which serves as the normalization.

\begin{figure}[ht]
\centering
\includegraphics[width=65mm]{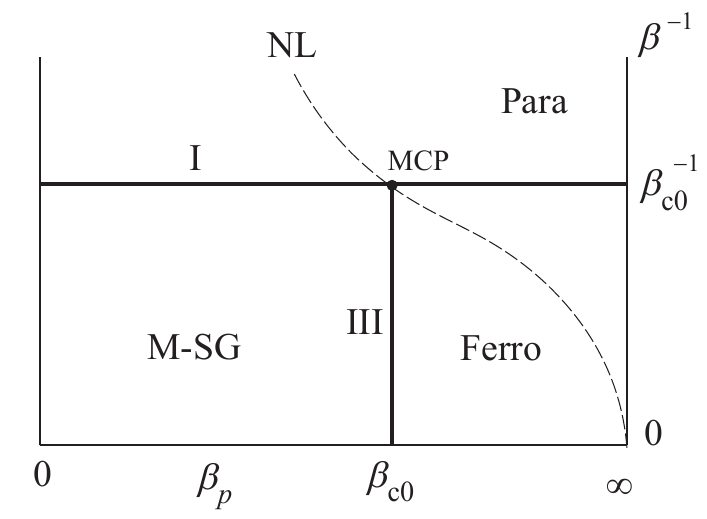}
\caption{Cross section of the phase diagram at $\gamma\to\infty$. $\beta_{c0}$ is the critical inverse temperature of the pure Ising model on the same lattice. The lower left part is the Mattis-like spin glass phase. NL is the Nishimori line $\beta=\beta_p$. Denoted by I is the boundary between the paramagnetic and ferromagnetic/spin glass phases, which is the cross section of the boundary I in the 3D phase diagram in Fig.~\ref{fig:3D_NoTC_NoReent}. The same applies to III as the boundary between the Mattis-like spin glass and ferromagnetic phases. MCP is the multicritical point.}
\label{fig:gamma_infinity}
\end{figure}

Consequently, the distribution function of the magnetization becomes
\begin{align}
   & P_1(x|\beta,\beta_p,\infty)\nonumber\\
    &=\sum_{\xi}\frac{e^{\beta_p \sum \xi_i\xi_j}}{Z_{\rm I}(\beta_p)}\frac{\sum_S\delta\big(x-\frac{1}{N}\sum_i S_i\big)e^{\beta\sum \xi_i\xi_j S_iS_j}}{Z_{\rm I}(\beta)}\nonumber\\
   &=\frac{1}{Z_{\rm I}(\beta_p)Z_{\rm I}(\beta)}\sum_{\xi,S}\delta
\big(x-\frac{1}{N}\sum_iS_i\big)e^{\beta_p \sum \xi_i\xi_j+\beta\sum \xi_i\xi_j S_iS_j}\nonumber\\
   &=\frac{1}{Z_{\rm I}(\beta_p)Z_{\rm I}(\beta)}\sum_{\xi,S}\delta
\big(x-\frac{1}{N}\sum_iS_i\xi_i\big)e^{\beta_p \sum \xi_i\xi_j+\beta\sum  S_iS_j}.
\label{eq:P1_gamma_infty}
\end{align}
  The above expression shows that this is the overlap of two independent pure Ising models with inverse temperatures $\beta_p$ and $\beta$. If we take the average of $x$, which is the magnetization, we find
\begin{align}
    m(\beta,\beta_p,\infty)=m_{\rm I}(\beta_p)m_{\rm I}(\beta),
\end{align}
the product of pure Ising magnetization values at $\beta_p$ and $\beta$. This quantity is non-zero if and only if $\beta_p>\beta_{\rm c0}$ and $\beta>\beta_{\rm c0}$, where $\beta_{\rm c0}$ is the critical inverse temperature of the pure Ising model on the given lattice. Therefore, the cross section of the 3D phase diagram for the limit $\gamma\to\infty$ is as shown in Fig. \ref{fig:gamma_infinity}. 

The lower left part of Fig.~\ref{fig:gamma_infinity} ($\beta^{-1}<{\beta_{\rm c0}}^{-1}$, $\beta_p<\beta_{\rm c0}$) is in the Mattis-like spin glass phase without frustration. 
The reason is that the magnetization vanishes in this region, but the spin glass order parameter $q$ is independent of $\beta_p$ and has the same value as in the case of $\beta_p=0$. The latter was discussed in Sec.~\ref{subsec:beta_p=0}, where it was argued that $q>0$ for $\beta^{-1}<\beta_{\rm c0}^{-1}$. 
This Mattis-like spin glass phase in the limit $\gamma\to\infty$ continues to the region with sufficiently large $\gamma$ as drawn in Fig.~\ref{fig:beta_p=0}.

It may be helpful to note that the left axis at $\beta_p = 0$ in Fig.~\ref{fig:gamma_infinity} corresponds to the right axis $\gamma \to \infty$ in Fig.~\ref{fig:beta_p=0}. The critical point at $\beta = \beta_{\rm c0}$ in the limit $\gamma \to \infty$ in Fig.~\ref{fig:beta_p=0} then extends horizontally in Fig.~\ref{fig:gamma_infinity}, marking the boundary between the paramagnetic phase and the ordered (Mattis-like spin glass or ferromagnetic) phase. This is because the spin-glass order parameter $q$, being gauge invariant, does not depend on $\beta_p$. Consequently, the boundary between the $q = 0$ (paramagnetic) and $q \ne 0$ (spin glass or ferromagnetic) phases appears as a horizontal line in the $\beta_p$-$\beta$ phase diagram.

It is also worth pointing out that there exists another type of the NL, defined by $\beta = \beta_p$ in the $\beta_p$-$\beta$ phase diagram for fixed $\gamma$, including the limit $\gamma \to \infty$, as will be discussed in the following subsection.

\subsection{$\gamma=0$}
\label{sub:gamma=0}
The model in the limit $\gamma = 0$ was examined in detail in Ref.~\cite{Nishimori2024}. For completeness, we briefly summarize the main results here, along with some generalizations to the case of small but finite $\gamma$.

The structure of the phase diagram depends on the presence or absence of temperature chaos in the Edwards-Anderson model. We therefore treat these two cases separately.

\subsubsection{No temperature chaos in the Edwards-Anderson model}
\label{subsub:noTC}
\begin{figure}[ht]
\centering
\includegraphics[width=65mm]{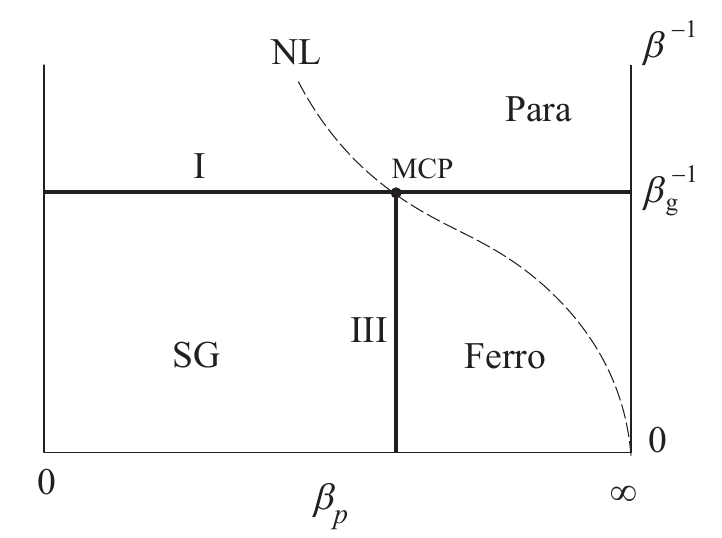}
\caption{Cross section of the phase diagram at $\gamma=0$ when temperature chaos does not exist in the Edwards-Anderson model. MCP is the multicritical point.  Denoted by I is the boundary between the paramagnetic and ferromagnetic/spin glass phases, which is the cross section of the boundary I in the 3D phase diagram in Fig.~\ref{fig:3D_NoTC_NoReent}. The same applies to III as the boundary between the spin glass and ferromagnetic phases. $\beta_{\rm g}^{-1}$ is the spin glass transition temperature of the Edwards-Anderson model.}
\label{fig:gamma=0 vertical}
\end{figure}

If there is no temperature chaos in the Edwards-Anderson model, the phase diagram for $\gamma=0$ with axes $\beta_p$ and $\beta$ has straight horizontal and vertical phase boundaries, see Fig.~\ref{fig:gamma=0 vertical}, as we show below.
The existence of the spin glass phase is assumed in the Edwards-Anderson model. Otherwise, there is no ordered phase at finite temperatures in this phase diagram, which is expected to be the case in two spatial dimensions \cite{Amoruso2004,Toldin2009,Ohzeki2009,Thomas2011}.

\vspace{2mm}
\noindent
\textit{Horizontal boundary}

The reason for the horizontal boundary, named I in Fig.~\ref{fig:gamma=0 vertical} separating the paramagnetic and ordered (spin glass/ferromagnetic) phases, is as follows.  We first note that the case $\beta_p=0$ corresponds to the Edwards-Anderson model with equal probabilities for $\tau=\pm 1$. Hence, the vertical axis at $\beta_p=0$ represents the Edwards-Anderson model with the spin glass transition at finite temperature $\beta_{\rm g}^{-1}$ accompanying singularities in the spin glass order parameter $q$ and the distribution function $P_2(x|\beta_1,\beta_2,0)$. Since these gauge-invariant quantities do not depend on $\beta_p$, see Eq.~(\ref{eq:P2}), the singularity should persist for any $\beta_p$ at the same temperature, leading to the horizontal phase boundary.

\vspace{2mm}
\noindent
\textit{Vertical boundary}

Physically, reentrance implies that a less ordered phase appears at lower temperature. We show that such reentrance contradicts the identity to relate the magnetization distribution and the overlap distribution, Eq.~(\ref{eq:p1_p2}).

\begin{figure}[ht]
\centering
\includegraphics[width=70mm]{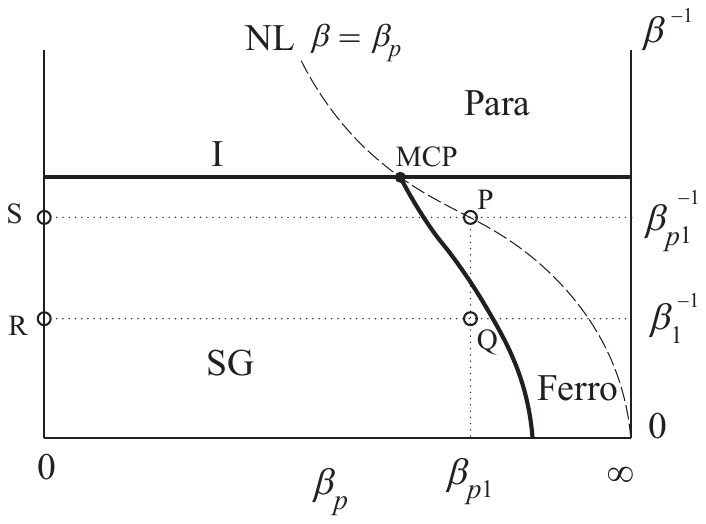}
\caption{Cross section of the phase diagram at $\gamma=0$ under the putative reentrant transition. It is shown in the main text that this structure is not allowed. Point P is on the NL ($\beta=\beta_p$) in the ferromagnetic phase and other points Q, S, and R are in the spin glass phase. Denoted by I is the boundary between the paramagnetic and ferromagnetic/spin glass phases.
}
\label{fig:gamma=0 reentrant}
\end{figure}

Suppose that reentrance exists, as illustrated in Fig.~\ref{fig:gamma=0 reentrant}, and consider point Q with coordinates Q$(\beta = \beta_1, \beta_p = \beta_{p1})$, which lies in the spin glass phase below the NL ($\beta = \beta_p$) and therefore $\beta_1 \ne \beta_{p1}$. The distribution function of the magnetization at point Q is a delta function centered at zero, reflecting the absence of spontaneous magnetization,
\begin{align}
    P_1(x|\beta_1, \beta_{p1}, 0) = \delta(x) \quad (\beta_1 \ne \beta_{p1}).
    \label{eq:P1_ne_delta}
\end{align}

Next, consider point P, which lies directly above Q on the NL. Point P has coordinates P$(\beta = \beta_{p1}, \beta_p = \beta_{p1})$ since the second coordinate is shared with Q by construction and the first coordinate is equal to the second on the NL.

According to Eq.~(\ref{eq:p1_p2}), the magnetization distribution at Q, $P_1(x|\beta_1, \beta_{p1}, 0)$, is equal to the overlap distribution $P_2(x|\beta_1, \beta_{p1}, 0)$, where the two spin configurations are sampled at inverse temperatures $\beta_1$ (point Q) and $\beta_{p1}$ (point P), respectively. As shown in Eq.~(\ref{eq:P2}), the overlap distribution $P_2(x|\beta_1, \beta_2, 0)$ is independent of $\beta_p$ and therefore constant along horizontal lines in Fig.~\ref{fig:gamma=0 reentrant}. Consequently, the overlap distribution between Q and P is equal to that between R and S. Therefore, we have
\begin{align}
    P_1(x|\beta_1, \beta_{p1}, 0) = P_2(x|\beta_1, \beta_{p1}, 0),
    \label{eq;p1=p2_gamma0}
\end{align}
where the left-hand side is the magnetization distribution at Q and the right-hand side is for the spin-state overlap between R and S.

Now recall that the vertical line at $\beta_p = 0$, where points R and S are located, corresponds to the Edwards-Anderson model. If the Edwards-Anderson model does not exhibit temperature chaos, the overlap distribution between R and S is nontrivial and cannot be a delta function:
\begin{align}
    P_2(x|\beta_1, \beta_{p1}, 0) \ne \delta(x) \quad (\beta_1 \ne \beta_{p1}), \label{eq:p2_ne_delta}
\end{align}
which contradicts Eqs.~(\ref{eq:P1_ne_delta}) and (\ref{eq;p1=p2_gamma0}). This contradiction implies that the reentrant phase diagram shown in Fig.~\ref{fig:gamma=0 reentrant} is not permissible under the assumption that the Edwards-Anderson model lacks temperature chaos.

A similar logic applies to the case with the ferromagnetic phase lying under the spin glass phase at lower temperatures. This proves that the phase boundary is vertical.

\vspace{2mm}
\noindent
\textit{Multicritical point on the NL}

\begin{figure}[ht]
\centering
\includegraphics[width=68mm]{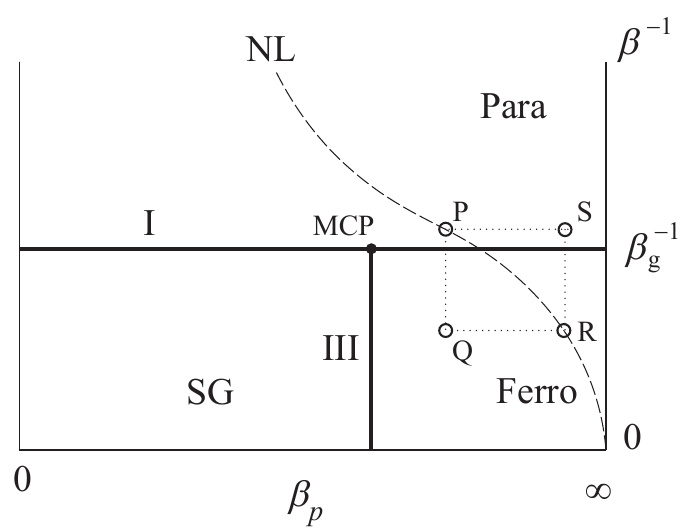}
\caption{Cross section of the phase diagram at $\gamma=0$ when the multicritical point is below the NL. It is shown in the main text that this structure is not allowed. Each point has the coordinate as: Q$(\beta_p,\beta)$, S$(\beta,\beta_p)$, P$(\beta_p,\beta_p)$, and R$(\beta,\beta)$. P and R are on the NL.}
\label{fig:gamma=0 MCP}
\end{figure}

We can show that the multicritical point, where three phases meet, lies on the NL $\beta=\beta_p$. Suppose that the multicritical point is below the NL as illustrated in Fig.~\ref{fig:gamma=0 MCP}. Points Q and S have coordinates with $\beta$ and $\beta_p$ exchanged, Q$(\beta_p,\beta)$ and S$(\beta,\beta_p)$. The reason is as follows. If P on the NL has $(\beta_p,\beta_p)$ and R on the NL has $(\beta,\beta)$, then the corresponding off-NL points are Q$(\beta_p,\beta)$ and S$(\beta,\beta_p)$. Then, Eq.~(\ref{eq:P1}) is not satisfied because Q is in the ferromagnetic phase and S is not.

Similarly, the NL does not lie below the multicritical point because, then, part of the NL is in the spin glass phase, which is not allowed by the identity on the NL $m(\beta,\beta,0)=q(\beta,\beta,0)$.

\vspace{2mm}
\noindent
\textit{Small but finite $\gamma$}

In the above analysis, we used Eqs.~(\ref{eq:p1_p2}) and (\ref{eq:P1}) with $\gamma = 0$. As seen in these equations, the logic developed above based on these equations remains valid for any finite $\gamma$. We also assumed the existence of a spin glass phase in the Edwards-Anderson model with equal probabilities for $\tau_{ij} = \pm 1$. This assumption is likely to hold even when the probabilities are unequal, i.e., for finite $\gamma$ in Eq.~(\ref{eq:P1}), as long as $\gamma$ remains below a certain threshold \cite{Ceccarelli2011}. Beyond this threshold, less frustrated configurations of $\tau$ become dominant with larger weights $Z_{\tau}(\gamma)$, and a different phase may emerge.

Consequently, the phase diagram shown in Fig.~\ref{fig:gamma=0 vertical} remains valid for finite values of $\gamma$ below a threshold. For larger $\gamma$, the spin glass phase is replaced by the less frustrated Mattis-like spin glass phase as discussed in Sec.~\ref{sub:mattis}. Even in this regime, the boundary between the ferromagnetic and Mattis-like spin glass phases remains a vertical straight line, according to Eq.~(\ref{eq:p1_p2}) as discussed above.

\vspace{2mm}
\noindent
\textit{Non-trivial distribution of the magnetization}

According to Eq.~(\ref{eq:p1_p2}) with $\beta = \beta_p$, the distribution function of the magnetization in the present model on the NL is equal to the distribution function of the replica overlap in the Edwards-Anderson model with disorder parameter $\gamma$. If replica symmetry breaking of the Parisi type \cite{Parisi1980} occurs in the latter, the distribution function $P_2(x|\beta,\beta,\gamma)$ takes a nontrivial form with support over a finite interval. Consequently, the magnetization distribution $P_1(x|\beta,\beta,\gamma)$ in the present model exhibits the same behavior. This implies that the magnetization can vary from one experimental realization to another, which is highly unusual for a macroscopic quantity.

Also derived is the possible existence of replica symmetry breaking on the NL $\beta=\beta_p$ since $P_2(x|\beta,\beta,\gamma)$ is nontrivial on the NL if it is non-trivial in the Edwards-Anderson model. If replica symmetry breaking indeed exists on the NL, it is in marked contrast to the case of uncorrelated disorder in the Edwards-Anderson model, where replica symmetry breaking is absent on the NL \cite{Nishimori2001b,Andrea2008,Barbier2022}.

\subsubsection{Temperature chaos in the Edwards-Anderson model}
\begin{figure}[ht]
\centering
\includegraphics[width=70mm]{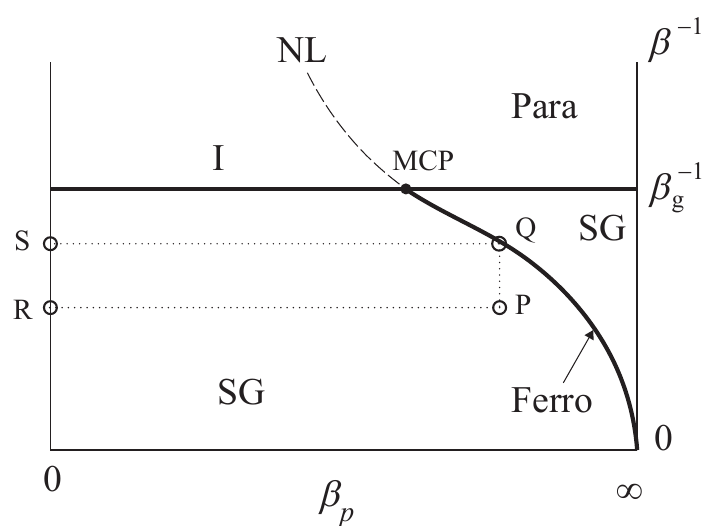}
\caption{Cross section of the phase diagram at $\gamma=0$ when temperature chaos exists in the Edwards-Anderson model. The coordinate of each point is:
P$(\beta_p,\beta)$, Q$(\beta_p,\beta_p)$,  R$(0,\beta)$,  S$(0,\beta_p)$.}
\label{fig:TC phase diagram}
\end{figure}

If temperature chaos exists in the Edwards-Anderson model, the ferromagnetic phase exists only on the NL as illustrated in Fig.~\ref{fig:TC phase diagram} because of Eq.~(\ref{eq:p1_p2}), which is reproduced here for the reader's convenience, 
\begin{align}
P_1(x|\beta,\beta_p,0)=P_2(x|\beta,\beta_p,0). \label{eq:p1_p2_2}
\end{align}
We apply this equation to point P in Fig.~\ref{fig:TC phase diagram} with the coordinate $(\beta_p,\beta)$, where $\beta_p\ne\beta$ since it is off the NL. Equation (\ref{eq:p1_p2_2}) implies that the distribution of the magnetization $P_1(x|\beta,\beta_p,0)$ at P is equal to the replica overlap of spin states $P_2(x|\beta,\beta_p,0)$ at R and S for the Edwards-Anderson model. The definition of temperature chaos is that the overlap of spin states vanishes when two states have different temperatures but with the same set of disorder variables,
\begin{align}
    P_2(x|\beta,\beta_p,0)=\delta (x)\quad (\beta\ne\beta_p).
\end{align}
Therefore, $P_1(x|\beta,\beta_p,0)$ for point P should also be a trivial delta function $\delta (x)$, demonstrating that there is no magnetization at P.

The same is true for finite $\gamma$ as long as it is below the threshold value, i.e., the corresponding Edwards-Anderson model is in the spin-glass phase as discussed before.

\subsection{3D phase diagram: Relation between temperature chaos and reentrance}
We proceed to integrate the findings from the previous subsections in order to construct the 3D phase diagram. This leads to the surprising result that temperature chaos and reentrance in the Edwards-Anderson model, previously regarded as unrelated phenomena, are intimately connected.

As in the preceding subsections, it is convenient to treat the cases with and without temperature chaos separately.

\subsubsection{No temperature chaos in the Edwards-Anderson model}
\begin{figure}[ht]
\centering
\includegraphics[width=82mm]{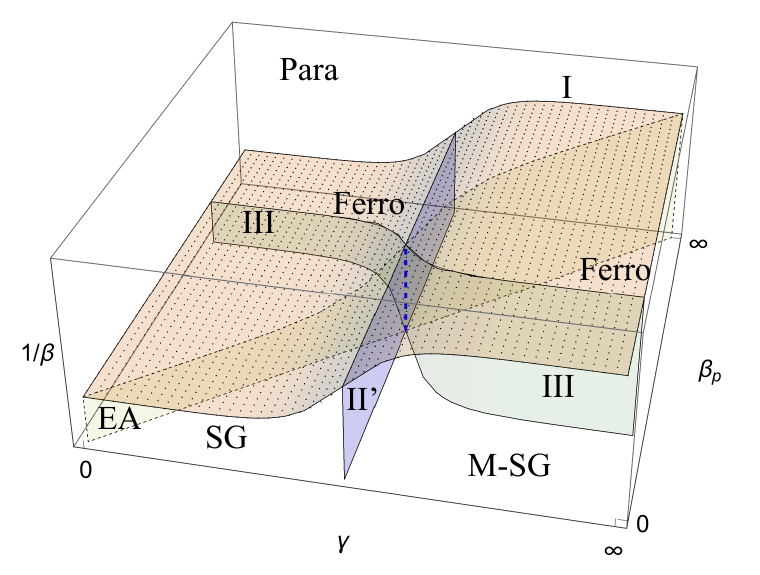}
\caption{3D phase diagram for the case without temperature chaos and reentrant transition in the Edwards-Anderson model. The existence of the spin glass phase is assumed.  Denoted by the symbols I, II', and III are the phase boundaries. The surface EA represents the Edwards-Anderson model $\gamma=\beta_p$, not a phase boundary. The ferromagnetic-spin glass boundary in the Edwards-Anderson model is a straight vertical line illustrated by the blue dashed line. Two ferromagnetic phases for large $\beta_p$ are separated by the boundary II' and have different characteristics.}
\label{fig:3D_NoTC_NoReent}
\end{figure}

The 3D phase diagram in the absence of temperature chaos but in the presence of the spin glass phase is shown in Fig.~\ref{fig:3D_NoTC_NoReent}.

To understand this structure, it is convenient to start from the boundary between the paramagnetic and non-paramagnetic (spin glass or ferromagnetic) phases, named I,  marking  the onset of the spin glass order parameter $q$, which is gauge invariant. The cross section with the plane $\beta_p=0$ is shown in Fig.~\ref{fig:beta_p=0}. As discussed in Sec.~\ref{subsec:beta_p=0}, this boundary I is an extension of the boundary between the paramagnetic and non-paramagnetic phases in the Edwards-Anderson model at $\beta_p=0$.
As mentioned in Sec.~\ref{subsec:beta_p_fixed}, this boundary has no dependence on $\beta_p$, and consequently the surface I for this boundary is flat along the $\beta_p$ axis in the 3D phase diagram. This fact is reflected in the straight horizontal boundary I in the constant-$\gamma$ cross sections in Fig.~\ref{fig:gamma_infinity} for $\gamma\to\infty$ and Fig.~\ref{fig:gamma=0 vertical} for $\gamma=0$.

Next is the boundary III between the ferromagnetic and spin glass phases. As shown in Sec. \ref{subsub:noTC}, this boundary is vertical for any constant-$\gamma$ cross section, as illustrated in Figs. \ref{fig:gamma_infinity} and \ref{fig:gamma=0 vertical}. This observation is reflected in the 3D phase diagram in Fig.~\ref{fig:3D_NoTC_NoReent} as the vertical surface III. This fact has a profound consequence, as explained next.

The vertical boundary III crosses the Edwards-Anderson plane $\gamma=\beta_p$ by a straight vertical, non-reentrant, line shown in blue dashed in Fig. ~\ref{fig:3D_NoTC_NoReent} marking the boundary between the ferromagnetic and spin glass phases in the Edwards-Anderson model. 
We have therefore derived the following Proposition. 
\begin{proposition}
    If the Edwards-Anderson model has a spin glass phase at finite temperature and the spin glass phase does not have temperature chaos, the boundary between the ferromagnetic and spin glass phases has no reentrance.
\end{proposition}

\noindent
{\bf Remark.}  Proposition 1 does not address what happens in reentrance if temperature chaos is present. In particular, reentrance may be absent when temperature chaos exists, as observed in the Sherrington–Kirkpatrick model \cite{Toulouse1980,GabayToulouse1981,Rizzo2003}. However, in finite dimensions, numerical evidence suggests that such a scenario is unlikely \cite{Nobre2001,Wang2003,Amoruso2004,Hasenbusch2007,Toldin2009,Ceccarelli2011,Thomas2011,Liu2025,Bray1987,Banavar1987,Fisher1986,Fisher1988, Kondor1989,Ney-Nifle_1997,Ney-Nifle1998,Parisi2010,Mathieu2001,Bouchaud2001,Aspelmeier2002,Rizzo2003,Houdayer2004,Katzgraber2007b,Fernandez_2013,Wang2015,Billoire2018,Baity-Jesi2021}.

\vspace{2mm}
This result is illustrated as the vertical boundary II' in Figs.~\ref{fig:beta_p=0} and \ref{fig:3D_NoTC_NoReent}.
There exist two ferromagnetic phases for large $\beta_p$ separated by the boundary II', one to the right of the plane II' for larger $\gamma$, and the other to the left for smaller $\gamma$. Both of these phases have finite magnetization $m\ne 0$ but have different characteristics. In the small-$\gamma$ ferromagnetic phase, the possible existence of support on a finite interval for the distribution function of the magnetization as discussed in Sec. \ref{subsub:noTC} suggests that this ferromagnetic phase may have some spin-glass characteristics. In contrast, the ferromagnetic phase with larger $\gamma$ shares the same gauge-invariant distribution of the overlap $P_2(x|\beta_1,\beta_2,\gamma)$ with the case $\beta_p=0$ in the Mattis-like spin glass phase.  The latter lacks intrinsic spin glass properties such as replica symmetry breaking. The boundary II' separates these different ferromagnetic phases.

\subsubsection{Temperature chaos in the Edwards-Anderson model}
We move on to the case with temperature chaos in the Edwards-Anderson model. To this end, it is useful to present Proposition 1 as its contrapositive. 

\begin{contrapositive*}
If the phase boundary between the ferromagnetic and non-ferromagnetic phases of the Edwards-Anderson model is reentrant, the model either exhibits temperature chaos in the spin glass phase or has no spin glass phase at finite temperature.
\end{contrapositive*}

\noindent
{\bf Remark 1.}  Proposition 1 and its contrapositive do not address what happens in temperature chaos if reentrance does not exist. Temperature chaos may or may not exist if there is no reentrance.

\vspace{2mm}

\noindent
{\bf Remark 2.}  Proposition 1 and its contrapositive establish a logical relationship between temperature chaos and reentrance. Nonetheless, the present theory does not determine the existence or absence of either phenomenon itself.

\vspace{2mm}

\begin{figure}[ht]
\centering
\includegraphics[width=75mm]{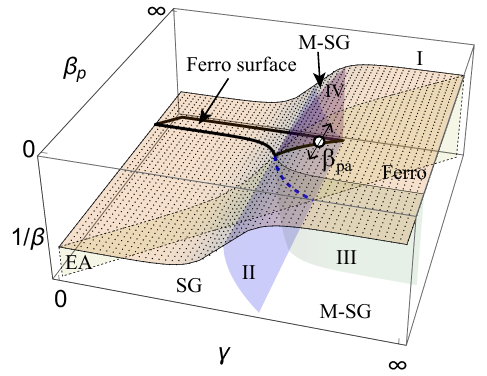}
\caption{3D phase diagram for the case with temperature chaos and reentrant transition in the Edwards-Anderson model. The existence of the spin glass phase is assumed.  Denoted by the symbols I, II,  III, and IV are the phase boundaries, the cross sections of which have been discussed in the previous subsection. The surface EA represents the Edwards-Anderson model $\gamma=\beta_p$, not a phase boundary. In the present case, the ferromagnetic-spin glass boundary in the Edwards-Anderson model is reentrant as illustrated by the blue dashed line, where EA and II cross. There is another ferromagnetic phase on the surface bordered by thick black lines.  Also, the Mattis-like spin glass phase exists in the region bordered by the plane II and the reentrant plane IV in the large-$\beta_p$ region. The point marked $\beta_{\rm pa}$ in the ferromagnetic phase moves horizontally as the point $\beta_{\rm pa}$ in Fig.~\ref{fig:EA2} moves vertically.}
\label{fig:3D_TC_Reent}
\end{figure}

Figure \ref{fig:3D_TC_Reent} is for the case with both reentrance and temperature chaos. 
When the spin glass phase in the Edwards-Anderson model has temperature chaos, the ferromagnetic phase in the region with large $\beta_p$ and small $\gamma$ is replaced by the spin glass phase except on the NL ($\beta_p=\beta$) as shown in Fig.~\ref{fig:TC phase diagram} for $\gamma=0$. In the 3D phase diagram, the ferromagnetic phase under NL condition $\beta_p=\beta$ exists on the surface surrounded by the thick black boundaries in Fig.~\ref{fig:3D_TC_Reent}.

It is worth calling attention to the following remarkable fact.
\begin{observation*}
The boundary III (between the ferromagnetic phase and the Mattis-like spin glass phase) has the same shape as the boundary I (between the ferromagnetic and paramagnetic phases). Similarly, the reentrant boundary II (between the Mattis-like spin glass phase and the spin glass phase) has the same shape as another reentrant boundary IV (between the ferromagnetic phase and the Mattis-like spin glass phase in the large-$\beta_p$ region). See Fig.~\ref{fig:3D_TC_Reent}.
\end{observation*}

To prove this Observation, let us consider the distribution function of the magnetization $P_1(x|\beta,\beta_p,\gamma)$ in the ferromagnetic phase in the region of large values $\gamma$, $\beta_p$, and $\beta$, the point marked $\beta_{\rm pa}$ in Fig.~\ref{fig:3D_TC_Reent}. We fix the values of $\beta$ to $\beta_{\rm a}$ and $\beta_p$ to $\beta_{\rm pa}$. In the identity of Eq.~(\ref{eq:p1_p2}),
\begin{align}
P_1(x|\beta_{\rm a},\beta_{\rm pa},\gamma)=P_2(x|\beta_{\rm a},\beta_{\rm pa},\gamma),
\end{align}
the right-hand side is the distribution function of the overlap for the Edwards-Anderson model with the disorder parameter $\gamma$.  This equation relates the property of the present model on the left-hand side $P_1(x|\beta_{\rm a},\beta_{\rm pa},\gamma)$ with the property of the Edwards-Anderson model on the right-hand side $P_2(x|\beta_{\rm a},\beta_{\rm pa},\gamma)$.

\begin{figure}[ht]
\centering
\includegraphics[width=65mm]{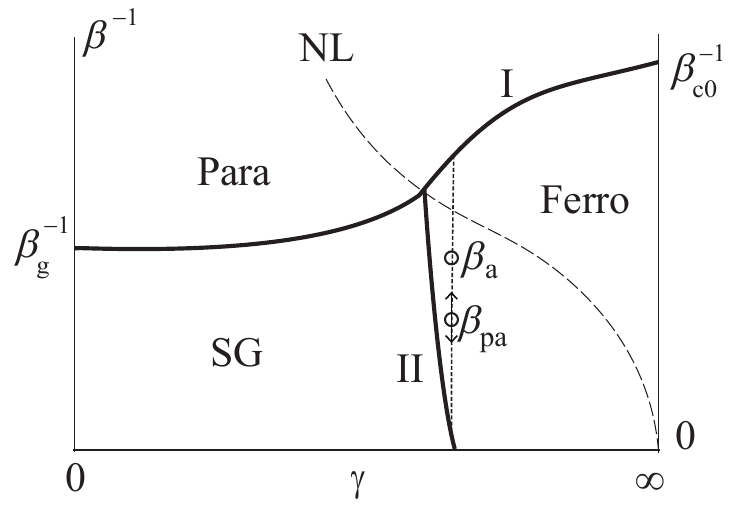}
\caption{Phase diagram of the original Edwards-Anderson model with the i.i.d. disorder variables under the parameter $\gamma$. Notice that this is different from Fig.~\ref{fig:beta_p=0}, where the large-$\gamma$ region is the Mattis-like spin glass phase. We consider two points marked $\beta_{\rm a}$ and $\beta_{\rm pa}$ and move the latter upward or downward along the vertical line shown dotted until it hits boundary I or boundary II. The corresponding point $\beta_{\rm pa}$ in the 3D phase diagram of Figs.~\ref{fig:3D_TC_Reent} and \ref{fig:2D_cut_3D_TC_Reent} moves horizontally until it hits the boundary III or boundary IV.} 
\label{fig:EA2}
\end{figure}

Now, on the right-hand side $P_2(x|\beta_{\rm a},\beta_{\rm pa},\gamma)$, we fix $\gamma$ to the range where the ferromagnetic phase exists at finite temperature and choose $\beta_{\rm a}$ and $\beta_{\rm pa}$ both lying in the ferromagnetic phase as illustrated in Fig.~\ref{fig:EA2}. Then, since no temperature chaos is expected to exist in the ferromagnetic phase, the right-hand side of the above equation has two delta peaks at the value of the spin-state overlap for inverse temperatures $\beta_{\rm a}$ and $\beta_{\rm pa}$,
\begin{align}
    P_2(x|\beta_{\rm a},\beta_{\rm pa},\gamma)=\frac{1}{2}\delta (x-q_{12})+\frac{1}{2}\delta (x+q_{12}).
\end{align}
Correspondingly, the left-hand side $P_1(x|\beta_{\rm a},\beta_{\rm pa},\gamma)$ for our present model has the same expression. This means that the point $(\beta_{\rm a},\beta_{\rm pa},\gamma)$ in the 3D phase diagram is in the ferromagnetic phase, as expected.

\begin{figure}[ht]
\centering
\includegraphics[width=68mm]{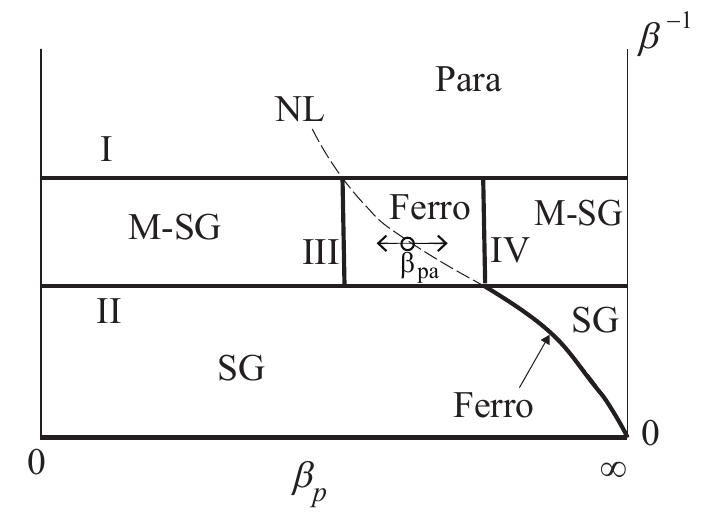}
\caption{2D cross section of Fig.~\ref{fig:3D_TC_Reent} with $\gamma$ fixed in the reentrant region. Boundaries I and II do not depend on $\beta_p$ and are horizontal lines whereas boundaries III and IV are independent of $\beta$ and thus appear as vertical lines. This structure is compatible with Eq.~(\ref{eq:P1}), which requires invariance of the magnetization distribution under the exchange of $\beta$ and $\beta_p$, as discussed in Sec.~\ref{subsub:noTC} on the location of the multicritical point using Fig.~\ref{fig:gamma=0 MCP}. The point marked $\beta_{\rm pa}$, corresponding to the point with the same symbol in Fig.~\ref{fig:3D_TC_Reent}, moves horizontally as the point $\beta_{\rm pa}$ in Fig.~\ref{fig:EA2} moves vertically.}
\label{fig:2D_cut_3D_TC_Reent}
\end{figure}

We next fix $\beta_{\rm a}$ and $\gamma$ and change $\beta_{\rm pa}$ from the initial value to smaller or larger values. Then, the point marked $\beta_{\rm pa}$ in Fig.~\ref{fig:EA2} moves upward or downward along the vertical line with fixed $\gamma$. When $\beta_{\rm pa}$ hits the para-ferro boundary I or the reentrant boundary II, $P_2(x|\beta_{\rm a},\beta_{\rm pa},\gamma)$ becomes trivial, $\delta(x)$. Correspondingly, the point $(\beta_{\rm a},\beta_{\rm pa},\gamma)$ in the 3D phase diagram (marked $\beta_{\rm pa}$ in Figs.~\ref{fig:3D_TC_Reent} and \ref{fig:2D_cut_3D_TC_Reent}) moves horizontally with $\gamma$ and $\beta=\beta_{\rm a}$ fixed, and hits the boundary III or boundary IV at exactly the same value of $\beta_{\rm pa}$ as the point $\beta_{\rm pa}$ hits I or II in Fig.~\ref{fig:EA2}. This holds true for any $\gamma$. Therefore, the shapes of the boundaries I and II in Fig.~\ref{fig:EA2} are precisely copied to those of III and IV in the 3D phase diagram. \qed

We have established that if there is reentrance in the Edwards-Anderson model as the boundary II in Fig.~\ref{fig:EA2}, our model also has a reentrant boundary IV. As $\beta_p$ increases, this boundary IV bends away from boundary II, as shown in Fig.~\ref{fig:3D_TC_Reent}. Then, there appears a region without magnetization between II and IV, which is another spin glass phase because the spin glass order parameter $q$ does not depend on $\beta_p$ and thus has the same value as in the Mattis-like spin glass phase at $\beta_p=0$. This region between II and IV is indeed a Mattis-like spin glass phase according to the discussion in Sec.~\ref{subsec:beta_p=0}.
A constant-$\gamma$ 2D cross section of  Fig.~\ref{fig:3D_TC_Reent} in the reentrant region is shown in Fig.~\ref{fig:2D_cut_3D_TC_Reent}. 

If there is no spin glass phase in the Edwards-Anderson model as expected in two spatial dimensions, the region to the left of the surface II (the small-$\gamma$ region) is occupied by the paramagnetic phase.

\subsection{Kitatani's model}
\begin{figure}[ht]
\centering
\includegraphics[width=70mm]{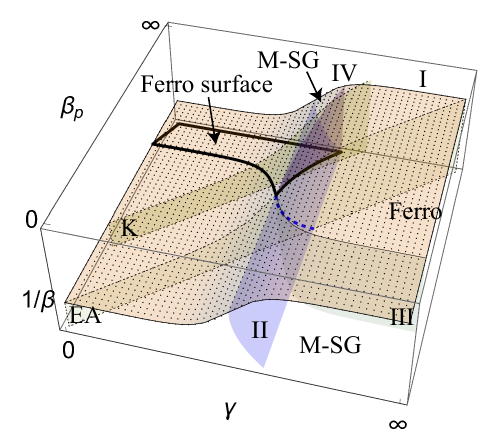}
\caption{Kitatani's model is denoted as K, a surface  shifted from the Edwards-Anderson model by the amount $a$. Part of this surface lies in the Mattis-like spin glass phase for large $\beta_p$ between II and IV if there is reentrance in the Edwards-Anderson model.}
\label{fig:kitatani}
\end{figure}

Kitatani introduced a model closely related to ours with a different parametrization corresponding to the replacement $\gamma\to K_p$ and $\beta_p\to K_p+a$ \cite{Kitatani1992}.  His model with $\beta_p$ shifted by $a$ from that of the Edwards-Anderson model ($\gamma=\beta_p$) is represented by the surface denoted K in Fig.~\ref{fig:kitatani}. He compared this model with the Edwards-Anderson model ($a=0$) and discussed that the latter does not have reentrance under the assumption that the ordered phase  above the NL of his model with $a>0$ is ferromagnetic.

As observed in Fig.~\ref{fig:kitatani}, part of his model is in the Mattis-like spin glass phase above the NL if there is reentrance in the Edwards-Anderson model. See also Fig.~\ref{fig:2D_cut_3D_TC_Reent}. This conflicts with his assumption that the ordered phase above the NL is always ferromagnetic. Conversely, if there is no reentrance in the Edwards-Anderson model, his assumption is valid and the absence of reentrance follows from his argument. However, this amounts to a tautology.

\subsection{Generalization}
Our theory applies also to the case with Gaussian disorder.  One simply replaces $\tau_{ij}$ by $J_{ij}$, and the summation over $\tau$ variables is replaced by the integral over $J_{ij}$,
\begin{align}
    \frac{1}{2^{N_{\rm B}}}\sum_{\tau}(\cdots) \longrightarrow \int_{-\infty}^{\infty} (\cdots)\prod_{\langle ij\rangle} \frac{e^{-\frac{{J_{ij}}^2}{2}}}{\sqrt{2\pi}}\, dJ_{ij}. \label{eq:gaussian}
\end{align}
The conclusions in the above sections remain valid in the Gaussian distribution.

Also, the $Z_q$ gauge glass  can be treated very similarly. Its Hamiltonian is \cite{Ozeki1993}, 
\begin{align}
    H=-\beta\sum_{\langle ij\rangle} \cos (\theta_i-\theta_j-\psi_{ij}),
\end{align}
where all angle variables can take one of the $q$ discrete values, $k/q~(k=0,1,\cdots, q-1)$ with the integer $q(\ge 2)$ fixed. The Ising model corresponds to $q=2$, and the model reduces to the XY gauge glass in the limit $q\to\infty$.

The gauge-invariant Potts model  also belongs to the class of models that can be analyzed in the same way. Following Ref.~\cite{Nishimori1983}, the Hamiltonian is
\begin{align}
    H=-\frac{\beta}{q}\sum_{\langle ij\rangle} \sum_{r=0}^{q-1}J_{ij}^{(r)}\lambda_i^r \lambda_j^{q-r},
\end{align}
where $\lambda_i=\omega^{k_i}$, $\omega=e^{2\pi i/q}$, and $k_i=0, 1,\cdots, q-1~(q\ge 2)$. The  disorder variable $J_{ij}^{(r)}=\tau_{ij}^{r}$ obeys the following distribution: $\tau_{ij}=1$ with probability $p$ and $\tau_{ij}$ being one of $\omega, \omega^2,\cdots, \omega^{q-1}$ with probability $(p-1)/(q-1)$ each. The Ising model is reproduced by $q=2$. 

All these Hamiltonians are gauge invariant, and the theory developed in the previous subsections applies with minimal adjustments.

\section{Conclusion}
The problem of spin glasses continues to be actively studied more than 50 years after its formulation as the Edwards-Anderson model~\cite{Edwards1975}. Its scope has since expanded well beyond the traditional boundaries of statistical physics~\cite{Nishimori2001,Mezard2009,Charbonneau2023,Zdeborova2016}. Despite this progress, only a limited number of analytical results with firm theoretical grounds have been established for the Edwards-Anderson model, particularly in two and three dimensions, which are the cases central to understanding experimental observations in real spin glass materials~\cite{Mydosh1993}. Insights based on solid analytical results are important not only for deepening theoretical understanding but also for providing independent verifications of numerical studies.

This paper takes a step toward a more comprehensive understanding of the Edwards-Anderson model through exact/rigorous analysis, by uncovering an unexpected connection between temperature chaos and reentrance, phenomena that are seemingly unrelated from a physical standpoint.
The central result of this work is that, assuming that a spin glass phase exists, {\em the presence of reentrance necessarily leads to the existence of temperature chaos in the Edwards-Anderson model}. This connection, though physically highly nontrivial, arises naturally through a symmetry-based analysis of our model with correlated disorder.
It is important to emphasize that Proposition~1 does not rule out the possibility of temperature chaos in the absence of reentrance, as stressed in the Remark to Proposition 1. This latter scenario applies, for instance, to the infinite-range Sherrington-Kirkpatrick model~\cite{Rizzo2003}. While we are unable to prove or disprove the existence of either temperature chaos or reentrance, our results have revealed a common nontrivial mathematical structure underlying these phenomena. The introduction of an additional degree of freedom in the disorder distribution has enabled us to identify this structure and to disclose the hidden connection between temperature chaos and reentrance through the 3D phase diagram.

A key strength of our approach lies in the modest analytical effort it entails. The argument relies on a small number of symmetry properties derived from the gauge symmetry inherent in the model, yet it demonstrates that the two complex physical phenomena, conventionally viewed as unrelated, are in fact closely connected. Similar symmetry-based arguments have previously yielded a range of nontrivial results for the Edwards-Anderson model~\cite{Nishimori1981,Nishimori1980,Nishimori1981thesis,Nishimori2001}, which, combined with the recent studies on correlated disorder~\cite{Nishimori2024,Nishimori2025}, provided  motivation for the present work. We hope that our approach serves as a fertile ground for further developments, both analytically and numerically, in the studies of the Edwards-Anderson model and related problems in the field of spin glasses and beyond.

\begin{acknowledgments}
Ohzeki and Okuyama were supported by JSPS KAKENHI Grant Nos. 24K16973 and 23H01432.
Ohzeki received financial supports by programs for Bridging the gap between R\&D and IDeal society (Society 5.0) and Generating Economic and social value (BRIDGE) and Cross-ministerial Strategic Innovation Promotion Program (SIP) from the Cabinet Office (No. 23836436).
\end{acknowledgments}

\begin{thebibliography}{58}%
\makeatletter
\providecommand \@ifxundefined [1]{%
 \@ifx{#1\undefined}
}%
\providecommand \@ifnum [1]{%
 \ifnum #1\expandafter \@firstoftwo
 \else \expandafter \@secondoftwo
 \fi
}%
\providecommand \@ifx [1]{%
 \ifx #1\expandafter \@firstoftwo
 \else \expandafter \@secondoftwo
 \fi
}%
\providecommand \natexlab [1]{#1}%
\providecommand \enquote  [1]{``#1''}%
\providecommand \bibnamefont  [1]{#1}%
\providecommand \bibfnamefont [1]{#1}%
\providecommand \citenamefont [1]{#1}%
\providecommand \href@noop [0]{\@secondoftwo}%
\providecommand \href [0]{\begingroup \@sanitize@url \@href}%
\providecommand \@href[1]{\@@startlink{#1}\@@href}%
\providecommand \@@href[1]{\endgroup#1\@@endlink}%
\providecommand \@sanitize@url [0]{\catcode `\\12\catcode `\$12\catcode `\&12\catcode `\#12\catcode `\^12\catcode `\_12\catcode `\%12\relax}%
\providecommand \@@startlink[1]{}%
\providecommand \@@endlink[0]{}%
\providecommand \url  [0]{\begingroup\@sanitize@url \@url }%
\providecommand \@url [1]{\endgroup\@href {#1}{\urlprefix }}%
\providecommand \urlprefix  [0]{URL }%
\providecommand \Eprint [0]{\href }%
\providecommand \doibase [0]{https://doi.org/}%
\providecommand \selectlanguage [0]{\@gobble}%
\providecommand \bibinfo  [0]{\@secondoftwo}%
\providecommand \bibfield  [0]{\@secondoftwo}%
\providecommand \translation [1]{[#1]}%
\providecommand \BibitemOpen [0]{}%
\providecommand \bibitemStop [0]{}%
\providecommand \bibitemNoStop [0]{.\EOS\space}%
\providecommand \EOS [0]{\spacefactor3000\relax}%
\providecommand \BibitemShut  [1]{\csname bibitem#1\endcsname}%
\let\auto@bib@innerbib\@empty
\bibitem [{\citenamefont {Sherrington}\ and\ \citenamefont {Kirkpatrick}(1975)}]{Sherrington1975}%
  \BibitemOpen
  \bibfield  {author} {\bibinfo {author} {\bibfnamefont {D.}~\bibnamefont {Sherrington}}\ and\ \bibinfo {author} {\bibfnamefont {S.}~\bibnamefont {Kirkpatrick}},\ }\bibfield  {title} {\bibinfo {title} {{Solvable model of a spin glass}},\ }\href {https://doi.org/10.1103/PhysRevLett.35.1792} {\bibfield  {journal} {\bibinfo  {journal} {Phys. Rev. Lett.}\ }\textbf {\bibinfo {volume} {35}},\ \bibinfo {pages} {1792} (\bibinfo {year} {1975})}\BibitemShut {NoStop}%
\bibitem [{\citenamefont {Parisi}(1980)}]{Parisi1980}%
  \BibitemOpen
  \bibfield  {author} {\bibinfo {author} {\bibfnamefont {G.}~\bibnamefont {Parisi}},\ }\bibfield  {title} {\bibinfo {title} {A sequence of approximated solutions to the {SK} model for spin glasses},\ }\href {https://doi.org/10.1088/0305-4470/13/4/009} {\bibfield  {journal} {\bibinfo  {journal} {J. Phys. A}\ }\textbf {\bibinfo {volume} {13}},\ \bibinfo {pages} {L115} (\bibinfo {year} {1980})}\BibitemShut {NoStop}%
\bibitem [{\citenamefont {Talagrand}(2006)}]{Talagrand2006}%
  \BibitemOpen
  \bibfield  {author} {\bibinfo {author} {\bibfnamefont {M.}~\bibnamefont {Talagrand}},\ }\bibfield  {title} {\bibinfo {title} {{The Parisi formula}},\ }\href {https://doi.org/10.4007/annals.2006.163.221} {\bibfield  {journal} {\bibinfo  {journal} {Ann. Math.}\ }\textbf {\bibinfo {volume} {163}},\ \bibinfo {pages} {221} (\bibinfo {year} {2006})}\BibitemShut {NoStop}%
\bibitem [{\citenamefont {M\'ezard}\ and\ \citenamefont {Montanari}(2009)}]{Mezard2009}%
  \BibitemOpen
  \bibfield  {author} {\bibinfo {author} {\bibfnamefont {M.}~\bibnamefont {M\'ezard}}\ and\ \bibinfo {author} {\bibfnamefont {A.}~\bibnamefont {Montanari}},\ }\href@noop {} {\emph {\bibinfo {title} {{Information, Physics, and Computation}}}}\ (\bibinfo  {publisher} {Oxford University Press},\ \bibinfo {address} {Oxford},\ \bibinfo {year} {2009})\BibitemShut {NoStop}%
\bibitem [{\citenamefont {Zdeborov{\'{a}}}\ and\ \citenamefont {Krzakala}(0116)}]{Zdeborova2016}%
  \BibitemOpen
  \bibfield  {author} {\bibinfo {author} {\bibfnamefont {L.}~\bibnamefont {Zdeborov{\'{a}}}}\ and\ \bibinfo {author} {\bibfnamefont {F.}~\bibnamefont {Krzakala}},\ }\bibfield  {title} {\bibinfo {title} {{Statistical physics of inference: thresholds and algorithms}},\ }\href {https://doi.org/10.1080/00018732.2016.1211393} {\bibfield  {journal} {\bibinfo  {journal} {Adv. Phys.}\ }\textbf {\bibinfo {volume} {65}},\ \bibinfo {pages} {453} (\bibinfo {year} {20116})}\BibitemShut {NoStop}%
\bibitem [{\citenamefont {Nishimori}(2001)}]{Nishimori2001}%
  \BibitemOpen
  \bibfield  {author} {\bibinfo {author} {\bibfnamefont {H.}~\bibnamefont {Nishimori}},\ }\href@noop {} {\emph {\bibinfo {title} {{Statistical Physics of Spin Glasses and Information Processing: An Introduction}}}}\ (\bibinfo  {publisher} {Oxford University Press},\ \bibinfo {address} {Oxford},\ \bibinfo {year} {2001})\BibitemShut {NoStop}%
\bibitem [{\citenamefont {Charbonneau}\ \emph {et~al.}(2023)\citenamefont {Charbonneau}, \citenamefont {Marinari}, \citenamefont {Parisi}, \citenamefont {Ricci-tersenghi}, \citenamefont {Sicuro}, \citenamefont {Zamponi},\ and\ \citenamefont {M\'ezard}}]{Charbonneau2023}%
  \BibitemOpen
  \bibfield  {author} {\bibinfo {author} {\bibfnamefont {P.}~\bibnamefont {Charbonneau}}, \bibinfo {author} {\bibfnamefont {E.}~\bibnamefont {Marinari}}, \bibinfo {author} {\bibfnamefont {G.}~\bibnamefont {Parisi}}, \bibinfo {author} {\bibfnamefont {F.}~\bibnamefont {Ricci-tersenghi}}, \bibinfo {author} {\bibfnamefont {G.}~\bibnamefont {Sicuro}}, \bibinfo {author} {\bibfnamefont {F.}~\bibnamefont {Zamponi}},\ and\ \bibinfo {author} {\bibfnamefont {M.}~\bibnamefont {M\'ezard}},\ }\href@noop {} {\emph {\bibinfo {title} {Spin Glass Theory and Far Beyond: Replica Symmetry Breaking after 40 Years}}}\ (\bibinfo  {publisher} {World Scientific},\ \bibinfo {address} {Singapore},\ \bibinfo {year} {2023})\BibitemShut {NoStop}%
\bibitem [{\citenamefont {Edwards}\ and\ \citenamefont {Anderson}(1975)}]{Edwards1975}%
  \BibitemOpen
  \bibfield  {author} {\bibinfo {author} {\bibfnamefont {S.~F.}\ \bibnamefont {Edwards}}\ and\ \bibinfo {author} {\bibfnamefont {P.~W.}\ \bibnamefont {Anderson}},\ }\bibfield  {title} {\bibinfo {title} {Theory of spin glasses},\ }\href {https://doi.org/10.1088/0305-4608/5/5/017} {\bibfield  {journal} {\bibinfo  {journal} {J. Phys. F}\ }\textbf {\bibinfo {volume} {5}},\ \bibinfo {pages} {965} (\bibinfo {year} {1975})}\BibitemShut {NoStop}%
\bibitem [{\citenamefont {Mydosh}(1993)}]{Mydosh1993}%
  \BibitemOpen
  \bibfield  {author} {\bibinfo {author} {\bibfnamefont {J.~A.}\ \bibnamefont {Mydosh}},\ }\href@noop {} {\emph {\bibinfo {title} {Spin glasses: An experimental introduction}}}\ (\bibinfo  {publisher} {CRC Press},\ \bibinfo {address} {London},\ \bibinfo {year} {1993})\BibitemShut {NoStop}%
\bibitem [{\citenamefont {Nobre}(2001)}]{Nobre2001}%
  \BibitemOpen
  \bibfield  {author} {\bibinfo {author} {\bibfnamefont {F.~D.}\ \bibnamefont {Nobre}},\ }\bibfield  {title} {\bibinfo {title} {Phase diagram of the two-dimensional {$\pm J$ Ising} spin glass},\ }\href {https://doi.org/10.1103/PhysRevE.64.046108} {\bibfield  {journal} {\bibinfo  {journal} {Phys. Rev. E}\ }\textbf {\bibinfo {volume} {64}},\ \bibinfo {pages} {046108} (\bibinfo {year} {2001})}\BibitemShut {NoStop}%
\bibitem [{\citenamefont {Wang}\ \emph {et~al.}(2003)\citenamefont {Wang}, \citenamefont {Harrington},\ and\ \citenamefont {Preskill}}]{Wang2003}%
  \BibitemOpen
  \bibfield  {author} {\bibinfo {author} {\bibfnamefont {C.}~\bibnamefont {Wang}}, \bibinfo {author} {\bibfnamefont {J.}~\bibnamefont {Harrington}},\ and\ \bibinfo {author} {\bibfnamefont {J.}~\bibnamefont {Preskill}},\ }\bibfield  {title} {\bibinfo {title} {Confinement-higgs transition in a disordered gauge theory and the accuracy threshold for quantum memory},\ }\href {https://doi.org/10.1016/S0003-4916(02)00019-2} {\bibfield  {journal} {\bibinfo  {journal} {Ann. Phys.}\ }\textbf {\bibinfo {volume} {303}},\ \bibinfo {pages} {31} (\bibinfo {year} {2003})}\BibitemShut {NoStop}%
\bibitem [{\citenamefont {Amoruso}\ and\ \citenamefont {Hartmann}(2004)}]{Amoruso2004}%
  \BibitemOpen
  \bibfield  {author} {\bibinfo {author} {\bibfnamefont {C.}~\bibnamefont {Amoruso}}\ and\ \bibinfo {author} {\bibfnamefont {A.~K.}\ \bibnamefont {Hartmann}},\ }\bibfield  {title} {\bibinfo {title} {Domain-wall energies and magnetization of the two-dimensional random-bond {Ising} model},\ }\href {https://doi.org/10.1103/PhysRevB.70.134425} {\bibfield  {journal} {\bibinfo  {journal} {Phys. Rev. B}\ }\textbf {\bibinfo {volume} {70}},\ \bibinfo {pages} {134425} (\bibinfo {year} {2004})}\BibitemShut {NoStop}%
\bibitem [{\citenamefont {Hasenbusch}\ \emph {et~al.}(2007)\citenamefont {Hasenbusch}, \citenamefont {Pelissetto},\ and\ \citenamefont {Vicari}}]{Hasenbusch2007}%
  \BibitemOpen
  \bibfield  {author} {\bibinfo {author} {\bibfnamefont {M.}~\bibnamefont {Hasenbusch}}, \bibinfo {author} {\bibfnamefont {A.}~\bibnamefont {Pelissetto}},\ and\ \bibinfo {author} {\bibfnamefont {E.}~\bibnamefont {Vicari}},\ }\bibfield  {title} {\bibinfo {title} {Critical behavior of three-dimensional {Ising} spin glass models},\ }\href {https://doi.org/10.1103/PhysRevB.76.094402} {\bibfield  {journal} {\bibinfo  {journal} {Phys. Rev. B}\ }\textbf {\bibinfo {volume} {76}},\ \bibinfo {pages} {094402} (\bibinfo {year} {2007})}\BibitemShut {NoStop}%
\bibitem [{\citenamefont {Toldin}\ \emph {et~al.}(2009)\citenamefont {Toldin}, \citenamefont {Pelissetto},\ and\ \citenamefont {Vicari}}]{Toldin2009}%
  \BibitemOpen
  \bibfield  {author} {\bibinfo {author} {\bibfnamefont {F.~P.}\ \bibnamefont {Toldin}}, \bibinfo {author} {\bibfnamefont {A.}~\bibnamefont {Pelissetto}},\ and\ \bibinfo {author} {\bibfnamefont {E.}~\bibnamefont {Vicari}},\ }\bibfield  {title} {\bibinfo {title} {Strong-disorder paramagnetic-ferromagnetic fixed point in the square-lattice {$\pm J$ Ising} model},\ }\href {https://doi.org/10.1007/s10955-009-9705-5} {\bibfield  {journal} {\bibinfo  {journal} {J. Stat. Phys.}\ }\textbf {\bibinfo {volume} {135}},\ \bibinfo {pages} {1039} (\bibinfo {year} {2009})}\BibitemShut {NoStop}%
\bibitem [{\citenamefont {Ceccarelli}\ \emph {et~al.}(2011)\citenamefont {Ceccarelli}, \citenamefont {Pelissetto},\ and\ \citenamefont {Vicari}}]{Ceccarelli2011}%
  \BibitemOpen
  \bibfield  {author} {\bibinfo {author} {\bibfnamefont {G.}~\bibnamefont {Ceccarelli}}, \bibinfo {author} {\bibfnamefont {A.}~\bibnamefont {Pelissetto}},\ and\ \bibinfo {author} {\bibfnamefont {E.}~\bibnamefont {Vicari}},\ }\bibfield  {title} {\bibinfo {title} {{Ferromagnetic-glassy transitions in three-dimensional Ising spin glasses}},\ }\href {https://doi.org/10.1103/PhysRevB.84.134202} {\bibfield  {journal} {\bibinfo  {journal} {Phys. Rev. B}\ }\textbf {\bibinfo {volume} {84}},\ \bibinfo {pages} {134202} (\bibinfo {year} {2011})}\BibitemShut {NoStop}%
\bibitem [{\citenamefont {Thomas}\ and\ \citenamefont {Katzgraber}(2011)}]{Thomas2011}%
  \BibitemOpen
  \bibfield  {author} {\bibinfo {author} {\bibfnamefont {C.~K.}\ \bibnamefont {Thomas}}\ and\ \bibinfo {author} {\bibfnamefont {H.~G.}\ \bibnamefont {Katzgraber}},\ }\bibfield  {title} {\bibinfo {title} {Simplest model to study reentrance in physical systems},\ }\href {https://doi.org/10.1103/PhysRevE.84.040101} {\bibfield  {journal} {\bibinfo  {journal} {Phys. Rev. E}\ }\textbf {\bibinfo {volume} {84}},\ \bibinfo {pages} {040101(R)} (\bibinfo {year} {2011})}\BibitemShut {NoStop}%
\bibitem [{\citenamefont {Liu}\ \emph {et~al.}(2025)\citenamefont {Liu}, \citenamefont {Wang}, \citenamefont {Wang}, \citenamefont {Yao},\ and\ \citenamefont {Tang}}]{Liu2025}%
  \BibitemOpen
  \bibfield  {author} {\bibinfo {author} {\bibfnamefont {Y.}~\bibnamefont {Liu}}, \bibinfo {author} {\bibfnamefont {D.}~\bibnamefont {Wang}}, \bibinfo {author} {\bibfnamefont {X.}~\bibnamefont {Wang}}, \bibinfo {author} {\bibfnamefont {D.-X.}\ \bibnamefont {Yao}},\ and\ \bibinfo {author} {\bibfnamefont {L.-H.}\ \bibnamefont {Tang}},\ }\bibfield  {title} {\bibinfo {title} {{Magnetization-resolved density of states and quasi-first order transition in the two-dimensional random bond Ising model: An entropic sampling study}},\ }\href {http://arxiv.org/abs/2505.04298} {\bibfield  {journal} {\bibinfo  {journal} {arXiv:2505.04298}\ } (\bibinfo {year} {2025})}\BibitemShut {NoStop}%
\bibitem [{\citenamefont {Bray}\ and\ \citenamefont {Moore}(1987)}]{Bray1987}%
  \BibitemOpen
  \bibfield  {author} {\bibinfo {author} {\bibfnamefont {A.~J.}\ \bibnamefont {Bray}}\ and\ \bibinfo {author} {\bibfnamefont {M.~A.}\ \bibnamefont {Moore}},\ }\bibfield  {title} {\bibinfo {title} {Chaotic nature of the spin-glass phase},\ }\href {https://doi.org/10.1103/PhysRevLett.58.57} {\bibfield  {journal} {\bibinfo  {journal} {Phys. Rev. Lett.}\ }\textbf {\bibinfo {volume} {58}},\ \bibinfo {pages} {57} (\bibinfo {year} {1987})}\BibitemShut {NoStop}%
\bibitem [{\citenamefont {Banavar}\ and\ \citenamefont {Bray}(1987)}]{Banavar1987}%
  \BibitemOpen
  \bibfield  {author} {\bibinfo {author} {\bibfnamefont {J.~R.}\ \bibnamefont {Banavar}}\ and\ \bibinfo {author} {\bibfnamefont {A.~J.}\ \bibnamefont {Bray}},\ }\bibfield  {title} {\bibinfo {title} {Chaos in spin glasses: A renormalization-group study},\ }\href {https://doi.org/10.1103/PhysRevB.35.8888} {\bibfield  {journal} {\bibinfo  {journal} {Phys. Rev. B}\ }\textbf {\bibinfo {volume} {35}},\ \bibinfo {pages} {8888} (\bibinfo {year} {1987})}\BibitemShut {NoStop}%
\bibitem [{\citenamefont {Fisher}\ and\ \citenamefont {Huse}(1986)}]{Fisher1986}%
  \BibitemOpen
  \bibfield  {author} {\bibinfo {author} {\bibfnamefont {D.~S.}\ \bibnamefont {Fisher}}\ and\ \bibinfo {author} {\bibfnamefont {D.~A.}\ \bibnamefont {Huse}},\ }\bibfield  {title} {\bibinfo {title} {Ordered phase of short-range {Ising} spin-glasses},\ }\href {https://doi.org/10.1103/PhysRevLett.56.1601} {\bibfield  {journal} {\bibinfo  {journal} {Phys. Rev. Lett.}\ }\textbf {\bibinfo {volume} {56}},\ \bibinfo {pages} {1601} (\bibinfo {year} {1986})}\BibitemShut {NoStop}%
\bibitem [{\citenamefont {Fisher}\ and\ \citenamefont {Huse}(1988)}]{Fisher1988}%
  \BibitemOpen
  \bibfield  {author} {\bibinfo {author} {\bibfnamefont {D.~S.}\ \bibnamefont {Fisher}}\ and\ \bibinfo {author} {\bibfnamefont {D.~A.}\ \bibnamefont {Huse}},\ }\bibfield  {title} {\bibinfo {title} {{Equilibrium behavior of the spin-glass ordered phase}},\ }\href {https://doi.org/10.1103/PhysRevB.38.386} {\bibfield  {journal} {\bibinfo  {journal} {Phys. Rev. B}\ }\textbf {\bibinfo {volume} {38}},\ \bibinfo {pages} {386} (\bibinfo {year} {1988})}\BibitemShut {NoStop}%
\bibitem [{\citenamefont {Kondor}(1989)}]{Kondor1989}%
  \BibitemOpen
  \bibfield  {author} {\bibinfo {author} {\bibfnamefont {I.}~\bibnamefont {Kondor}},\ }\bibfield  {title} {\bibinfo {title} {On chaos in spin glasses},\ }\href {https://doi.org/10.1088/0305-4470/22/5/005} {\bibfield  {journal} {\bibinfo  {journal} {J. Phys. A}\ }\textbf {\bibinfo {volume} {22}},\ \bibinfo {pages} {L163} (\bibinfo {year} {1989})}\BibitemShut {NoStop}%
\bibitem [{\citenamefont {Ney-Nifle}\ and\ \citenamefont {Young}(1997)}]{Ney-Nifle-1997}%
  \BibitemOpen
  \bibfield  {author} {\bibinfo {author} {\bibfnamefont {M.}~\bibnamefont {Ney-Nifle}}\ and\ \bibinfo {author} {\bibfnamefont {A.~P.}\ \bibnamefont {Young}},\ }\bibfield  {title} {\bibinfo {title} {Chaos in a two-dimensional ising spin glass},\ }\href {https://doi.org/10.1088/0305-4470/30/15/017} {\bibfield  {journal} {\bibinfo  {journal} {J. Phys. A}\ }\textbf {\bibinfo {volume} {30}},\ \bibinfo {pages} {5311} (\bibinfo {year} {1997})}\BibitemShut {NoStop}%
\bibitem [{\citenamefont {Ney-Nifle}(1998)}]{Ney-Nifle1998}%
  \BibitemOpen
  \bibfield  {author} {\bibinfo {author} {\bibfnamefont {M.}~\bibnamefont {Ney-Nifle}},\ }\bibfield  {title} {\bibinfo {title} {Chaos and universality in a four-dimensional spin glass},\ }\href {https://doi.org/10.1103/PhysRevB.57.492} {\bibfield  {journal} {\bibinfo  {journal} {Phys. Rev. B}\ }\textbf {\bibinfo {volume} {57}},\ \bibinfo {pages} {492} (\bibinfo {year} {1998})}\BibitemShut {NoStop}%
\bibitem [{\citenamefont {Parisi}\ and\ \citenamefont {Rizzo}(2010)}]{Parisi2010}%
  \BibitemOpen
  \bibfield  {author} {\bibinfo {author} {\bibfnamefont {G.}~\bibnamefont {Parisi}}\ and\ \bibinfo {author} {\bibfnamefont {T.}~\bibnamefont {Rizzo}},\ }\bibfield  {title} {\bibinfo {title} {{Chaos in temperature in diluted mean-field spin-glass}},\ }\href {https://doi.org/10.1088/1751-8113/43/23/235003} {\bibfield  {journal} {\bibinfo  {journal} {J. Phys. A}\ }\textbf {\bibinfo {volume} {43}},\ \bibinfo {pages} {235003} (\bibinfo {year} {2010})}\BibitemShut {NoStop}%
\bibitem [{\citenamefont {Mathieu}\ \emph {et~al.}(2001)\citenamefont {Mathieu}, \citenamefont {J\"onsson}, \citenamefont {Nordblad}, \citenamefont {Katori},\ and\ \citenamefont {Ito}}]{Mathieu2001}%
  \BibitemOpen
  \bibfield  {author} {\bibinfo {author} {\bibfnamefont {R.}~\bibnamefont {Mathieu}}, \bibinfo {author} {\bibfnamefont {P.~E.}\ \bibnamefont {J\"onsson}}, \bibinfo {author} {\bibfnamefont {P.}~\bibnamefont {Nordblad}}, \bibinfo {author} {\bibfnamefont {H.~A.}\ \bibnamefont {Katori}},\ and\ \bibinfo {author} {\bibfnamefont {A.}~\bibnamefont {Ito}},\ }\bibfield  {title} {\bibinfo {title} {Memory and chaos in an {Ising} spin glass},\ }\href {https://doi.org/10.1103/PhysRevB.65.012411} {\bibfield  {journal} {\bibinfo  {journal} {Phys. Rev. B}\ }\textbf {\bibinfo {volume} {65}},\ \bibinfo {pages} {012411} (\bibinfo {year} {2001})}\BibitemShut {NoStop}%
\bibitem [{\citenamefont {Bouchaud}\ \emph {et~al.}(2001)\citenamefont {Bouchaud}, \citenamefont {Dupuis}, \citenamefont {Hammann},\ and\ \citenamefont {Vincent}}]{Bouchaud2001}%
  \BibitemOpen
  \bibfield  {author} {\bibinfo {author} {\bibfnamefont {J.-P.}\ \bibnamefont {Bouchaud}}, \bibinfo {author} {\bibfnamefont {V.}~\bibnamefont {Dupuis}}, \bibinfo {author} {\bibfnamefont {J.}~\bibnamefont {Hammann}},\ and\ \bibinfo {author} {\bibfnamefont {E.}~\bibnamefont {Vincent}},\ }\bibfield  {title} {\bibinfo {title} {Separation of time and length scales in spin glasses: Temperature as a microscope},\ }\href {https://doi.org/10.1103/PhysRevB.65.024439} {\bibfield  {journal} {\bibinfo  {journal} {Phys. Rev. B}\ }\textbf {\bibinfo {volume} {65}},\ \bibinfo {pages} {024439} (\bibinfo {year} {2001})}\BibitemShut {NoStop}%
\bibitem [{\citenamefont {Aspelmeier}\ \emph {et~al.}(2002)\citenamefont {Aspelmeier}, \citenamefont {Bray},\ and\ \citenamefont {Moore}}]{Aspelmeier2002}%
  \BibitemOpen
  \bibfield  {author} {\bibinfo {author} {\bibfnamefont {T.}~\bibnamefont {Aspelmeier}}, \bibinfo {author} {\bibfnamefont {A.~J.}\ \bibnamefont {Bray}},\ and\ \bibinfo {author} {\bibfnamefont {M.~A.}\ \bibnamefont {Moore}},\ }\bibfield  {title} {\bibinfo {title} {Why temperature chaos in spin glasses is hard to observe},\ }\href {https://doi.org/10.1103/PhysRevLett.89.197202} {\bibfield  {journal} {\bibinfo  {journal} {Phys. Rev. Lett.}\ }\textbf {\bibinfo {volume} {89}},\ \bibinfo {pages} {197202} (\bibinfo {year} {2002})}\BibitemShut {NoStop}%
\bibitem [{\citenamefont {Rizzo}\ and\ \citenamefont {Crisanti}(2003)}]{Rizzo2003}%
  \BibitemOpen
  \bibfield  {author} {\bibinfo {author} {\bibfnamefont {T.}~\bibnamefont {Rizzo}}\ and\ \bibinfo {author} {\bibfnamefont {A.}~\bibnamefont {Crisanti}},\ }\bibfield  {title} {\bibinfo {title} {{Chaos in temperature in the Sherrington-Kirkpatrick model}},\ }\href {https://doi.org/10.1103/PhysRevLett.90.137201} {\bibfield  {journal} {\bibinfo  {journal} {Phys. Rev. Lett.}\ }\textbf {\bibinfo {volume} {90}},\ \bibinfo {pages} {4} (\bibinfo {year} {2003})}\BibitemShut {NoStop}%
\bibitem [{\citenamefont {Houdayer}\ and\ \citenamefont {Hartmann}(2004)}]{Houdayer2004}%
  \BibitemOpen
  \bibfield  {author} {\bibinfo {author} {\bibfnamefont {J.}~\bibnamefont {Houdayer}}\ and\ \bibinfo {author} {\bibfnamefont {A.~K.}\ \bibnamefont {Hartmann}},\ }\bibfield  {title} {\bibinfo {title} {Low-temperature behavior of two-dimensional gaussian {Ising} spin glasses},\ }\href {https://doi.org/10.1103/PhysRevB.70.014418} {\bibfield  {journal} {\bibinfo  {journal} {Phys. Rev. B}\ }\textbf {\bibinfo {volume} {70}},\ \bibinfo {pages} {014418} (\bibinfo {year} {2004})}\BibitemShut {NoStop}%
\bibitem [{\citenamefont {Katzgraber}\ and\ \citenamefont {Krza\ifmmode~\mbox{\c{}}\else \c{}\fi{}ka\l{}a}(2007)}]{Katzgraber2007b}%
  \BibitemOpen
  \bibfield  {author} {\bibinfo {author} {\bibfnamefont {H.~G.}\ \bibnamefont {Katzgraber}}\ and\ \bibinfo {author} {\bibfnamefont {F.}~\bibnamefont {Krza\ifmmode~\mbox{\c{}}\else \c{}\fi{}ka\l{}a}},\ }\bibfield  {title} {\bibinfo {title} {Temperature and disorder chaos in three-dimensional {Ising} spin glasses},\ }\href {https://doi.org/10.1103/PhysRevLett.98.017201} {\bibfield  {journal} {\bibinfo  {journal} {Phys. Rev. Lett.}\ }\textbf {\bibinfo {volume} {98}},\ \bibinfo {pages} {017201} (\bibinfo {year} {2007})}\BibitemShut {NoStop}%
\bibitem [{\citenamefont {Fernandez}\ \emph {et~al.}(2013)\citenamefont {Fernandez}, \citenamefont {Martin-Mayor}, \citenamefont {Parisi},\ and\ \citenamefont {Seoane}}]{Fernandez-2013}%
  \BibitemOpen
  \bibfield  {author} {\bibinfo {author} {\bibfnamefont {L.~A.}\ \bibnamefont {Fernandez}}, \bibinfo {author} {\bibfnamefont {V.}~\bibnamefont {Martin-Mayor}}, \bibinfo {author} {\bibfnamefont {G.}~\bibnamefont {Parisi}},\ and\ \bibinfo {author} {\bibfnamefont {B.}~\bibnamefont {Seoane}},\ }\bibfield  {title} {\bibinfo {title} {Temperature chaos in 3d {Ising} spin glasses is driven by rare events},\ }\href {https://doi.org/10.1209/0295-5075/103/67003} {\bibfield  {journal} {\bibinfo  {journal} {Europhys. Lett.}\ }\textbf {\bibinfo {volume} {103}},\ \bibinfo {pages} {67003} (\bibinfo {year} {2013})}\BibitemShut {NoStop}%
\bibitem [{\citenamefont {Wang}\ \emph {et~al.}(2015)\citenamefont {Wang}, \citenamefont {Machta},\ and\ \citenamefont {Katzgraber}}]{Wang2015}%
  \BibitemOpen
  \bibfield  {author} {\bibinfo {author} {\bibfnamefont {W.}~\bibnamefont {Wang}}, \bibinfo {author} {\bibfnamefont {J.}~\bibnamefont {Machta}},\ and\ \bibinfo {author} {\bibfnamefont {H.~G.}\ \bibnamefont {Katzgraber}},\ }\bibfield  {title} {\bibinfo {title} {Chaos in spin glasses revealed through thermal boundary conditions},\ }\href {https://doi.org/10.1103/PhysRevB.92.094410} {\bibfield  {journal} {\bibinfo  {journal} {Phys. Rev. B}\ }\textbf {\bibinfo {volume} {92}},\ \bibinfo {pages} {094410} (\bibinfo {year} {2015})}\BibitemShut {NoStop}%
\bibitem [{\citenamefont {Billoire}\ \emph {et~al.}(2018)\citenamefont {Billoire}, \citenamefont {Fernandez}, \citenamefont {Maiorano}, \citenamefont {Marinari}, \citenamefont {Martin-Mayor}, \citenamefont {Moreno-Gordo}, \citenamefont {Parisi}, \citenamefont {Ricci-Tersenghi},\ and\ \citenamefont {Ruiz-Lorenzo}}]{Billoire2018}%
  \BibitemOpen
  \bibfield  {author} {\bibinfo {author} {\bibfnamefont {A.}~\bibnamefont {Billoire}}, \bibinfo {author} {\bibfnamefont {L.~A.}\ \bibnamefont {Fernandez}}, \bibinfo {author} {\bibfnamefont {A.}~\bibnamefont {Maiorano}}, \bibinfo {author} {\bibfnamefont {E.}~\bibnamefont {Marinari}}, \bibinfo {author} {\bibfnamefont {V.}~\bibnamefont {Martin-Mayor}}, \bibinfo {author} {\bibfnamefont {J.}~\bibnamefont {Moreno-Gordo}}, \bibinfo {author} {\bibfnamefont {G.}~\bibnamefont {Parisi}}, \bibinfo {author} {\bibfnamefont {F.}~\bibnamefont {Ricci-Tersenghi}},\ and\ \bibinfo {author} {\bibfnamefont {J.~J.}\ \bibnamefont {Ruiz-Lorenzo}},\ }\bibfield  {title} {\bibinfo {title} {{Dynamic variational study of chaos: Spin glasses in three dimensions}},\ }\href {https://doi.org/10.1088/1742-5468/aaa387} {\bibfield  {journal} {\bibinfo  {journal} {J. Stat. Mech.}\ }\textbf {\bibinfo {volume} {2018}},\ \bibinfo {pages} {033302} (\bibinfo {year} {2018})}\BibitemShut {NoStop}%
\bibitem [{\citenamefont {Baity-Jesi}\ \emph {et~al.}(2021)\citenamefont {Baity-Jesi}, \citenamefont {Calore}, \citenamefont {Cruz}, \citenamefont {Fernandez}, \citenamefont {Gil-Narvion}, \citenamefont {{Gonzalez-Adalid Pemartin}}, \citenamefont {Gordillo-Guerrero}, \citenamefont {I{\~{n}}iguez}, \citenamefont {Maiorano}, \citenamefont {Marinari}, \citenamefont {Martin-Mayor}, \citenamefont {Moreno-Gordo}, \citenamefont {Mu{\~{n}}oz-Sudupe}, \citenamefont {Navarro}, \citenamefont {Paga}, \citenamefont {Parisi}, \citenamefont {Perez-Gaviro}, \citenamefont {Ricci-Tersenghi}, \citenamefont {Ruiz-Lorenzo}, \citenamefont {Schifano}, \citenamefont {Seoane}, \citenamefont {Tarancon}, \citenamefont {Tripiccione},\ and\ \citenamefont {Yllanes}}]{Baity-Jesi2021}%
  \BibitemOpen
  \bibfield  {author} {\bibinfo {author} {\bibfnamefont {M.}~\bibnamefont {Baity-Jesi}}, \bibinfo {author} {\bibfnamefont {E.}~\bibnamefont {Calore}}, \bibinfo {author} {\bibfnamefont {A.}~\bibnamefont {Cruz}}, \bibinfo {author} {\bibfnamefont {L.~A.}\ \bibnamefont {Fernandez}}, \bibinfo {author} {\bibfnamefont {J.~M.}\ \bibnamefont {Gil-Narvion}}, \bibinfo {author} {\bibfnamefont {I.}~\bibnamefont {{Gonzalez-Adalid Pemartin}}}, \bibinfo {author} {\bibfnamefont {A.}~\bibnamefont {Gordillo-Guerrero}}, \bibinfo {author} {\bibfnamefont {D.}~\bibnamefont {I{\~{n}}iguez}}, \bibinfo {author} {\bibfnamefont {A.}~\bibnamefont {Maiorano}}, \bibinfo {author} {\bibfnamefont {E.}~\bibnamefont {Marinari}}, \bibinfo {author} {\bibfnamefont {V.}~\bibnamefont {Martin-Mayor}}, \bibinfo {author} {\bibfnamefont {J.}~\bibnamefont {Moreno-Gordo}}, \bibinfo {author} {\bibfnamefont {A.}~\bibnamefont {Mu{\~{n}}oz-Sudupe}}, \bibinfo {author} {\bibfnamefont {D.}~\bibnamefont {Navarro}}, \bibinfo {author} {\bibfnamefont
  {I.}~\bibnamefont {Paga}}, \bibinfo {author} {\bibfnamefont {G.}~\bibnamefont {Parisi}}, \bibinfo {author} {\bibfnamefont {S.}~\bibnamefont {Perez-Gaviro}}, \bibinfo {author} {\bibfnamefont {F.}~\bibnamefont {Ricci-Tersenghi}}, \bibinfo {author} {\bibfnamefont {J.~J.}\ \bibnamefont {Ruiz-Lorenzo}}, \bibinfo {author} {\bibfnamefont {S.~F.}\ \bibnamefont {Schifano}}, \bibinfo {author} {\bibfnamefont {B.}~\bibnamefont {Seoane}}, \bibinfo {author} {\bibfnamefont {A.}~\bibnamefont {Tarancon}}, \bibinfo {author} {\bibfnamefont {R.}~\bibnamefont {Tripiccione}},\ and\ \bibinfo {author} {\bibfnamefont {D.}~\bibnamefont {Yllanes}},\ }\bibfield  {title} {\bibinfo {title} {{Temperature chaos is present in off-equilibrium spin-glass dynamics}},\ }\href {https://doi.org/10.1038/s42005-021-00565-9} {\bibfield  {journal} {\bibinfo  {journal} {Commun. Phys.}\ }\textbf {\bibinfo {volume} {4}},\ \bibinfo {pages} {74} (\bibinfo {year} {2021})}\BibitemShut {NoStop}%
\bibitem [{\citenamefont {Nishimori}(2024)}]{Nishimori2024}%
  \BibitemOpen
  \bibfield  {author} {\bibinfo {author} {\bibfnamefont {H.}~\bibnamefont {Nishimori}},\ }\bibfield  {title} {\bibinfo {title} {{Anomalous distribution of magnetization in an Ising spin glass with correlated disorder}},\ }\href {https://doi.org/10.1103/PhysRevE.110.064108} {\bibfield  {journal} {\bibinfo  {journal} {Phys. Rev. E}\ }\textbf {\bibinfo {volume} {110}},\ \bibinfo {pages} {064108} (\bibinfo {year} {2024})}\BibitemShut {NoStop}%
\bibitem [{\citenamefont {Nishimori}(2025)}]{Nishimori2025}%
  \BibitemOpen
  \bibfield  {author} {\bibinfo {author} {\bibfnamefont {H.}~\bibnamefont {Nishimori}},\ }\bibfield  {title} {\bibinfo {title} {{Instability of the ferromagnetic phase under random fields in an Ising spin glass with correlated disorder}},\ }\href {https://doi.org/10.1103/PhysRevE.111.044109} {\bibfield  {journal} {\bibinfo  {journal} {Phys. Rev. E}\ }\textbf {\bibinfo {volume} {111}},\ \bibinfo {pages} {044109} (\bibinfo {year} {2025})}\BibitemShut {NoStop}%
\bibitem [{\citenamefont {Braunstein}\ \emph {et~al.}(2025)\citenamefont {Braunstein}, \citenamefont {Budzynski}, \citenamefont {Mariani},\ and\ \citenamefont {Ricci-Tersenghi}}]{Braunstein2025}%
  \BibitemOpen
  \bibfield  {author} {\bibinfo {author} {\bibfnamefont {A.}~\bibnamefont {Braunstein}}, \bibinfo {author} {\bibfnamefont {L.}~\bibnamefont {Budzynski}}, \bibinfo {author} {\bibfnamefont {M.}~\bibnamefont {Mariani}},\ and\ \bibinfo {author} {\bibfnamefont {F.}~\bibnamefont {Ricci-Tersenghi}},\ }\bibfield  {title} {\bibinfo {title} {{Evidence of replica symmetry breaking under the Nishimori conditions in epidemic inference on graphs}},\ }\href {https://doi.org/10.1103/tbd5-25vf} {\bibfield  {journal} {\bibinfo  {journal} {Phys. Rev. E}\ }\textbf {\bibinfo {volume} {111}},\ \bibinfo {pages} {064308} (\bibinfo {year} {2025})}\BibitemShut {NoStop}%
\bibitem [{\citenamefont {Nishimori}(1981)}]{Nishimori1981}%
  \BibitemOpen
  \bibfield  {author} {\bibinfo {author} {\bibfnamefont {H.}~\bibnamefont {Nishimori}},\ }\bibfield  {title} {\bibinfo {title} {Internal energy, specific heat and correlation function of the bond-random {Ising} model},\ }\href {https://doi.org/10.1143/PTP.66.1169} {\bibfield  {journal} {\bibinfo  {journal} {Prog. Theor. Phys.}\ }\textbf {\bibinfo {volume} {66}},\ \bibinfo {pages} {1169} (\bibinfo {year} {1981})}\BibitemShut {NoStop}%
\bibitem [{\citenamefont {Nishimori}(1980)}]{Nishimori1980}%
  \BibitemOpen
  \bibfield  {author} {\bibinfo {author} {\bibfnamefont {H.}~\bibnamefont {Nishimori}},\ }\bibfield  {title} {\bibinfo {title} {Exact results and critical properties of the {Ising} model with competing interactions},\ }\href {https://doi.org/10.1088/0022-3719/13/21/012} {\bibfield  {journal} {\bibinfo  {journal} {J. Phys. C}\ }\textbf {\bibinfo {volume} {13}},\ \bibinfo {pages} {4071} (\bibinfo {year} {1980})}\BibitemShut {NoStop}%
\bibitem [{\citenamefont {Nishimori}(1982)}]{Nishimori1981thesis}%
  \BibitemOpen
  \bibfield  {author} {\bibinfo {author} {\bibfnamefont {H.}~\bibnamefont {Nishimori}},\ }\emph {\bibinfo {title} {Rigorous Results on Random Spin Systems with Competing Interactions}},\ \href@noop {} {Ph.D. thesis},\ \bibinfo  {school} {University of Tokyo} (\bibinfo {year} {1982})\BibitemShut {NoStop}%
\bibitem [{\citenamefont {Kitatani}(1992)}]{Kitatani1992}%
  \BibitemOpen
  \bibfield  {author} {\bibinfo {author} {\bibfnamefont {H.}~\bibnamefont {Kitatani}},\ }\bibfield  {title} {\bibinfo {title} {The verticality of the ferromagnetic-spin glass phase boundary of the {$\pm J$ Ising} model in the {$p$-$T$} plane},\ }\href {https://doi.org/10.1143/JPSJ.61.4049} {\bibfield  {journal} {\bibinfo  {journal} {J. Phys. Soc. Jpn.}\ }\textbf {\bibinfo {volume} {61}},\ \bibinfo {pages} {4049} (\bibinfo {year} {1992})}\BibitemShut {NoStop}%
\bibitem [{\citenamefont {Hasenbusch}\ \emph {et~al.}(2008)\citenamefont {Hasenbusch}, \citenamefont {Pelissetto},\ and\ \citenamefont {Vicari}}]{Hasenbusch2008}%
  \BibitemOpen
  \bibfield  {author} {\bibinfo {author} {\bibfnamefont {M.}~\bibnamefont {Hasenbusch}}, \bibinfo {author} {\bibfnamefont {A.}~\bibnamefont {Pelissetto}},\ and\ \bibinfo {author} {\bibfnamefont {E.}~\bibnamefont {Vicari}},\ }\bibfield  {title} {\bibinfo {title} {{Critical behavior of three-dimensional Ising spin glass models}},\ }\href {https://doi.org/10.1103/PhysRevB.78.214205} {\bibfield  {journal} {\bibinfo  {journal} {Phy. Rev. B}\ }\textbf {\bibinfo {volume} {78}},\ \bibinfo {pages} {214205} (\bibinfo {year} {2008})}\BibitemShut {NoStop}%
\bibitem [{\citenamefont {Altieri}\ and\ \citenamefont {Baity-Jesi}(2023)}]{AltieriBaityJesi2023}%
  \BibitemOpen
  \bibfield  {author} {\bibinfo {author} {\bibfnamefont {A.}~\bibnamefont {Altieri}}\ and\ \bibinfo {author} {\bibfnamefont {M.}~\bibnamefont {Baity-Jesi}},\ }\bibfield  {title} {\bibinfo {title} {An introduction to the theory of spin glasses},\ }in\ \href@noop {} {\emph {\bibinfo {booktitle} {Reference Module in Materials Science and Materials Engineering}}},\ \bibinfo {editor} {edited by\ \bibinfo {editor} {\bibfnamefont {F.~A.}\ \bibnamefont {Gozzo}}\ and\ \bibinfo {editor} {\bibfnamefont {Q.}~\bibnamefont {Jiang}}}\ (\bibinfo  {publisher} {Elsevier},\ \bibinfo {year} {2023})\ p.\ \bibinfo {pages} {361}\BibitemShut {NoStop}%
\bibitem [{\citenamefont {Dahlberg}\ \emph {et~al.}(2024)\citenamefont {Dahlberg}, \citenamefont {Pemart{\'{i}}n}, \citenamefont {Marinari}, \citenamefont {Martin-Mayor}, \citenamefont {Moreno-Gordo}, \citenamefont {Orbach}, \citenamefont {Paga}, \citenamefont {Parisi}, \citenamefont {Ricci-Tersenghi}, \citenamefont {Ruiz-Lorenzo},\ and\ \citenamefont {Yllanes}}]{Dahlberg2024}%
  \BibitemOpen
  \bibfield  {author} {\bibinfo {author} {\bibfnamefont {E.~D.}\ \bibnamefont {Dahlberg}}, \bibinfo {author} {\bibfnamefont {I.~G.-A.}\ \bibnamefont {Pemart{\'{i}}n}}, \bibinfo {author} {\bibfnamefont {E.}~\bibnamefont {Marinari}}, \bibinfo {author} {\bibfnamefont {V.}~\bibnamefont {Martin-Mayor}}, \bibinfo {author} {\bibfnamefont {J.}~\bibnamefont {Moreno-Gordo}}, \bibinfo {author} {\bibfnamefont {R.~L.}\ \bibnamefont {Orbach}}, \bibinfo {author} {\bibfnamefont {I.}~\bibnamefont {Paga}}, \bibinfo {author} {\bibfnamefont {G.}~\bibnamefont {Parisi}}, \bibinfo {author} {\bibfnamefont {F.}~\bibnamefont {Ricci-Tersenghi}}, \bibinfo {author} {\bibfnamefont {J.~J.}\ \bibnamefont {Ruiz-Lorenzo}},\ and\ \bibinfo {author} {\bibfnamefont {D.}~\bibnamefont {Yllanes}},\ }\bibfield  {title} {\bibinfo {title} {{Spin-glass dynamics: experiment, theory and simulation}},\ }\href {https://doi.org/10.48550/arXiv.2412.08381} {\bibfield  {journal} {\bibinfo  {journal} {arXiv:2412.08381}\ } (\bibinfo {year} {2024})}\BibitemShut
  {NoStop}%
\bibitem [{\citenamefont {Horiguchi}\ and\ \citenamefont {Morita}(1982)}]{Horiguchi1982}%
  \BibitemOpen
  \bibfield  {author} {\bibinfo {author} {\bibfnamefont {T.}~\bibnamefont {Horiguchi}}\ and\ \bibinfo {author} {\bibfnamefont {T.}~\bibnamefont {Morita}},\ }\bibfield  {title} {\bibinfo {title} {Existence of the ferromagnetic phase in a random-bond {Ising} model on the square lattice},\ }\href {https://doi.org/10.1088/0305-4470/15/2/005} {\bibfield  {journal} {\bibinfo  {journal} {J. Phys. A}\ }\textbf {\bibinfo {volume} {15}},\ \bibinfo {pages} {L75} (\bibinfo {year} {1982})}\BibitemShut {NoStop}%
\bibitem [{\citenamefont {Garban}\ and\ \citenamefont {Spencer}(2022)}]{garban2022continuous}%
  \BibitemOpen
  \bibfield  {author} {\bibinfo {author} {\bibfnamefont {C.}~\bibnamefont {Garban}}\ and\ \bibinfo {author} {\bibfnamefont {T.}~\bibnamefont {Spencer}},\ }\bibfield  {title} {\bibinfo {title} {Continuous symmetry breaking along the {Nishimori} line},\ }\href {https://doi.org/10.1063/5.0087024} {\bibfield  {journal} {\bibinfo  {journal} {J. Math. Phys.}\ }\textbf {\bibinfo {volume} {63}},\ \bibinfo {pages} {093302} (\bibinfo {year} {2022})}\BibitemShut {NoStop}%
\bibitem [{\citenamefont {Chen}\ \emph {et~al.}(2025)\citenamefont {Chen}, \citenamefont {Zhu}, \citenamefont {Verresen}, \citenamefont {Seif}, \citenamefont {B{\"{a}}umer}, \citenamefont {Layden}, \citenamefont {Tantivasadakarn}, \citenamefont {Zhu}, \citenamefont {Sheldon}, \citenamefont {Vishwanath}, \citenamefont {Trebst},\ and\ \citenamefont {Kandala}}]{Chen2025}%
  \BibitemOpen
  \bibfield  {author} {\bibinfo {author} {\bibfnamefont {E.~H.}\ \bibnamefont {Chen}}, \bibinfo {author} {\bibfnamefont {G.-Y.}\ \bibnamefont {Zhu}}, \bibinfo {author} {\bibfnamefont {R.}~\bibnamefont {Verresen}}, \bibinfo {author} {\bibfnamefont {A.}~\bibnamefont {Seif}}, \bibinfo {author} {\bibfnamefont {E.}~\bibnamefont {B{\"{a}}umer}}, \bibinfo {author} {\bibfnamefont {D.}~\bibnamefont {Layden}}, \bibinfo {author} {\bibfnamefont {N.}~\bibnamefont {Tantivasadakarn}}, \bibinfo {author} {\bibfnamefont {G.}~\bibnamefont {Zhu}}, \bibinfo {author} {\bibfnamefont {S.}~\bibnamefont {Sheldon}}, \bibinfo {author} {\bibfnamefont {A.}~\bibnamefont {Vishwanath}}, \bibinfo {author} {\bibfnamefont {S.}~\bibnamefont {Trebst}},\ and\ \bibinfo {author} {\bibfnamefont {A.}~\bibnamefont {Kandala}},\ }\bibfield  {title} {\bibinfo {title} {Nishimori transition across the error threshold for constant-depth quantum circuits},\ }\href {https://doi.org/10.1038/s41567-024-02696-6} {\bibfield  {journal} {\bibinfo  {journal} {Nature
  Phys.}\ }\textbf {\bibinfo {volume} {211}},\ \bibinfo {pages} {161} (\bibinfo {year} {2025})}\BibitemShut {NoStop}%
\bibitem [{\citenamefont {Zhu}\ \emph {et~al.}(2023)\citenamefont {Zhu}, \citenamefont {Tantivasadakarn}, \citenamefont {Vishwanath}, \citenamefont {Trebst},\ and\ \citenamefont {Verresen}}]{Zhu2023}%
  \BibitemOpen
  \bibfield  {author} {\bibinfo {author} {\bibfnamefont {G.-Y.}\ \bibnamefont {Zhu}}, \bibinfo {author} {\bibfnamefont {N.}~\bibnamefont {Tantivasadakarn}}, \bibinfo {author} {\bibfnamefont {A.}~\bibnamefont {Vishwanath}}, \bibinfo {author} {\bibfnamefont {S.}~\bibnamefont {Trebst}},\ and\ \bibinfo {author} {\bibfnamefont {R.}~\bibnamefont {Verresen}},\ }\bibfield  {title} {\bibinfo {title} {Nishimori's cat: Stable long-range entanglement from finite-depth unitaries and weak measurements},\ }\href {https://doi.org/10.1103/PhysRevLett.131.200201} {\bibfield  {journal} {\bibinfo  {journal} {Phys. Rev. Lett.}\ }\textbf {\bibinfo {volume} {131}},\ \bibinfo {pages} {200201} (\bibinfo {year} {2023})}\BibitemShut {NoStop}%
\bibitem [{\citenamefont {Mattis}(1976)}]{Mattis1976}%
  \BibitemOpen
  \bibfield  {author} {\bibinfo {author} {\bibfnamefont {D.~C.}\ \bibnamefont {Mattis}},\ }\bibfield  {title} {\bibinfo {title} {Solvable spin systems with random interactions},\ }\href {https://doi.org/10.1016/0375-9601(76)90396-0} {\bibfield  {journal} {\bibinfo  {journal} {Phys. Lett. A}\ }\textbf {\bibinfo {volume} {56}},\ \bibinfo {pages} {421} (\bibinfo {year} {1976})}\BibitemShut {NoStop}%
\bibitem [{\citenamefont {Ohzeki}\ and\ \citenamefont {Nishimori}(2009)}]{Ohzeki2009}%
  \BibitemOpen
  \bibfield  {author} {\bibinfo {author} {\bibfnamefont {M.}~\bibnamefont {Ohzeki}}\ and\ \bibinfo {author} {\bibfnamefont {H.}~\bibnamefont {Nishimori}},\ }\bibfield  {title} {\bibinfo {title} {{Analytical evidence for the absence of spin glass transition on self-dual lattices}},\ }\href {https://doi.org/10.1088/1751-8113/42/33/332001} {\bibfield  {journal} {\bibinfo  {journal} {J. Phys. A}\ }\textbf {\bibinfo {volume} {42}},\ \bibinfo {pages} {332001} (\bibinfo {year} {2009})}\BibitemShut {NoStop}%
\bibitem [{\citenamefont {Nishimori}\ and\ \citenamefont {Sherrington}(2001)}]{Nishimori2001b}%
  \BibitemOpen
  \bibfield  {author} {\bibinfo {author} {\bibfnamefont {H.}~\bibnamefont {Nishimori}}\ and\ \bibinfo {author} {\bibfnamefont {D.}~\bibnamefont {Sherrington}},\ }\bibfield  {title} {\bibinfo {title} {Absence of replica symmetry breaking in a region of the phase diagram of the {Ising} spin glass},\ }\href {https://doi.org/10.1063/1.1358165} {\bibfield  {journal} {\bibinfo  {journal} {AIP Conf. Proc.}\ }\textbf {\bibinfo {volume} {553}},\ \bibinfo {pages} {67} (\bibinfo {year} {2001})}\BibitemShut {NoStop}%
\bibitem [{\citenamefont {Andrea}(2008)}]{Andrea2008}%
  \BibitemOpen
  \bibfield  {author} {\bibinfo {author} {\bibfnamefont {M.}~\bibnamefont {Andrea}},\ }\bibfield  {title} {\bibinfo {title} {{Estimating random variables from random sparse observations}},\ }\href {https://doi.org/10.1002/ett.1289} {\bibfield  {journal} {\bibinfo  {journal} {Euro. Trans.Telecommun.}\ }\textbf {\bibinfo {volume} {19}},\ \bibinfo {pages} {385} (\bibinfo {year} {2008})}\BibitemShut {NoStop}%
\bibitem [{\citenamefont {Barbier}\ and\ \citenamefont {Panchenko}(2022)}]{Barbier2022}%
  \BibitemOpen
  \bibfield  {author} {\bibinfo {author} {\bibfnamefont {J.}~\bibnamefont {Barbier}}\ and\ \bibinfo {author} {\bibfnamefont {D.}~\bibnamefont {Panchenko}},\ }\bibfield  {title} {\bibinfo {title} {Strong replica symmetry in high-dimensional optimal {Bayesian} inference},\ }\href {https://doi.org/10.1007/s00220-022-04387-w} {\bibfield  {journal} {\bibinfo  {journal} {Commun. Math. Phys.}\ }\textbf {\bibinfo {volume} {393}},\ \bibinfo {pages} {1199} (\bibinfo {year} {2022})}\BibitemShut {NoStop}%
\bibitem [{\citenamefont {Toulouse}(1980)}]{Toulouse1980}%
  \BibitemOpen
  \bibfield  {author} {\bibinfo {author} {\bibfnamefont {G.}~\bibnamefont {Toulouse}},\ }\bibfield  {title} {\bibinfo {title} {On the mean field theory of mixed spin glass-ferromagnetic phases},\ }\href {https://doi.org/10.1051/jphyslet:019800041018044700} {\bibfield  {journal} {\bibinfo  {journal} {J. Phys. (Paris)}\ }\textbf {\bibinfo {volume} {41}},\ \bibinfo {pages} {447} (\bibinfo {year} {1980})}\BibitemShut {NoStop}%
\bibitem [{\citenamefont {Gabay}\ and\ \citenamefont {Toulouse}(1981)}]{GabayToulouse1981}%
  \BibitemOpen
  \bibfield  {author} {\bibinfo {author} {\bibfnamefont {M.}~\bibnamefont {Gabay}}\ and\ \bibinfo {author} {\bibfnamefont {G.}~\bibnamefont {Toulouse}},\ }\bibfield  {title} {\bibinfo {title} {Coexistence of spin-glass and ferromagnetic orderings},\ }\href {https://doi.org/10.1103/PhysRevLett.47.201} {\bibfield  {journal} {\bibinfo  {journal} {Phys. Rev. Lett.}\ }\textbf {\bibinfo {volume} {47}},\ \bibinfo {pages} {201} (\bibinfo {year} {1981})}\BibitemShut {NoStop}%
\bibitem [{\citenamefont {Ozeki}\ and\ \citenamefont {Nishimori}(1993)}]{Ozeki1993}%
  \BibitemOpen
  \bibfield  {author} {\bibinfo {author} {\bibfnamefont {Y.}~\bibnamefont {Ozeki}}\ and\ \bibinfo {author} {\bibfnamefont {H.}~\bibnamefont {Nishimori}},\ }\bibfield  {title} {\bibinfo {title} {{Phase diagram of gauge glasses}},\ }\href {https://doi.org/10.1088/0305-4470/26/14/009} {\bibfield  {journal} {\bibinfo  {journal} {J. Phys. A}\ }\textbf {\bibinfo {volume} {26}},\ \bibinfo {pages} {3399} (\bibinfo {year} {1993})}\BibitemShut {NoStop}%
\bibitem [{\citenamefont {Nishimori}\ and\ \citenamefont {Stephen}(1983)}]{Nishimori1983}%
  \BibitemOpen
  \bibfield  {author} {\bibinfo {author} {\bibfnamefont {H.}~\bibnamefont {Nishimori}}\ and\ \bibinfo {author} {\bibfnamefont {M.~J.}\ \bibnamefont {Stephen}},\ }\bibfield  {title} {\bibinfo {title} {{Gauge-invariant frustrated Potts spin glass}},\ }\href {https://doi.org/https://doi.org/10.1103/PhysRevB.27.5644} {\bibfield  {journal} {\bibinfo  {journal} {Phys. Rev. B}\ }\textbf {\bibinfo {volume} {27}},\ \bibinfo {pages} {5644} (\bibinfo {year} {1983})}\BibitemShut {NoStop}%
\end{thebibliography}
%

\end{document}